\documentclass[12pt,oneside, a4paper]{article}
\ifx\pdfoutput\undefined
\usepackage[dvips,bookmarks=false]{hyperref}	% This is for arXiv.org
\else
\usepackage{hyperref}	% This is for pdftex
\fi
\hypersetup{colorlinks,bookmarksopen,bookmarksnumbered,citecolor=blue,
linkcolor=black,pdfstartview=FitH,urlcolor=blue}

\usepackage{graphicx}
\usepackage{amssymb}
\usepackage{cite}
\usepackage{bm}
\usepackage{indentfirst}
\usepackage{amsmath}
\usepackage{hhline}
\usepackage{multirow}

\usepackage{ulem}
\begin{document}

\begin{titlepage}

\begin{flushright}
KANAZAWA-18-02\\
TUM-HEP/1142/18
\end{flushright}

\begin{center}

\vspace{1cm}
{\large\bf 
Boosted Self-interacting Dark Matter in a Multi-component
 Dark Matter Model}
\vspace{1cm}

\renewcommand{\thefootnote}{\fnsymbol{footnote}}
Mayumi Aoki$^1$\footnote[1]{mayumi@hep.s.kanazawa-u.ac.jp}
,
Takashi Toma$^{2}$\footnote[2]{takashi.toma@tum.de}
\vspace{5mm}

{\it%
$^1${Institute for Theoretical Physics, Kanazawa University, Kanazawa
 920-1192, Japan}
$^2${Physik-Department T30d, Technische Universit\"at M\"unchen,\\
 James-Franck-Stra\ss{}e, D-85748 Garching, Germany}
}

\vspace{8mm}

\abstract{
 In models of multi-component dark matter, the lighter component of dark
 matter can be
 boosted by annihilations of the heavier state if mass splitting is
 large enough. Such relativistic dark matter can be detectable via large
 neutrino detectors such as Super-Kamiokande and IceCube.
 Moreover, if the process is inelastic scattering and decay length of
 the produced particle is short enough, another signature
 coming from the decay can also be detectable. 
 In this paper, we construct a simple two-component dark matter model
 with a hidden $U(1)_D$ gauge symmetry where the lighter component of
 dark matter has a potential to improve 
 the so-called small scale structure problems with large
 self-interacting cross section. 
 We estimate number of multi-Cherenkov ring events due to both of the
 boosted dark matter 
 and subsequent decay of the particle produced by inelastic scattering
 at Hyper-Kamiokande future experiment.
 Some relevant constraints, such as dark matter direct detection and 
 cosmological observations, are also taken into account. 
 The numerical analysis shows that some parameter space which can
 induce large self-interacting cross section can give a few multi-Cherenkov ring
 events per year at Hyper-Kamiokande.
 }

\end{center}
\end{titlepage}

\renewcommand{\thefootnote}{\arabic{footnote}}
\newcommand{\bhline}[1]{\noalign{\hrule height #1}}
\newcommand{\bvline}[1]{\vrule width #1}

\setcounter{footnote}{0}

\setcounter{page}{1}
%%%%%%%%%%%%%%%%%%%%%%%%%%%%%%%%%%%%%%

\section{Introduction}
From cosmological observations, it is clear that dark matter exists in
the universe. 
However our knowledge about dark matter is limited.
We know that $26\%$ of the total energy density in the universe is
occupied by dark matter~\cite{Ade:2015xua}, but the nature of the
particle of dark matter, mass and interaction other than gravity are unknown.
Revealing the nature of dark 
matter is one of the most important issues in (astro-)particle physics. 

From the point of view of dark matter model building, many models have been
proposed so far, and in those models 
one component dark matter is often considered just for simplicity.
Exploring the simplest possibility at first would actually be the best option. 
However there is no strong motivation to consider only one component dark
matter. 
In fact, multi-component dark matter naturally appears and is
phenomenologically motivated in some
cases~\cite{Berezhiani:1989fp, Boehm:2003ha, Chialva:2012rq,
Aoki:2013gzs, Kajiyama:2013rla, Bhattacharya:2013hva, Esch:2014jpa, 
Aoki:2014lha, Karam:2016rsz, Bhattacharya:2016ysw, DiFranzo:2016uzc,
Arcadi:2016kmk, Borah:2017xgm, Bhattacharya:2017fid, Ahmed:2017dbb,
Aoki:2017eqn}, and some interesting 
consequences are expected such as double disk galaxy structure
constructed by normal matter and dark matter~\cite{Fan:2013yva,
Fan:2013tia}, multiple gamma-ray line 
signals and boosted dark matter signals~\cite{Agashe:2014yua,
Kong:2014mia, Alhazmi:2016qcs, Kim:2016zjx, Giudice:2017zke,
Chatterjee:2018mej, Kim:2018veo} and so on. 

In particular, boosted dark matter is an interesting consequence of
multi-component dark matter though it can also be
produced by dark matter semi-annihilations or decay of a 
long-lived particle for example~\cite{Agashe:2014yua, Kong:2014mia,
Alhazmi:2016qcs, Kim:2016zjx, Giudice:2017zke, Chatterjee:2018mej, Kim:2018veo}. 
If the heavier component of dark matter annihilates into the lighter dark
matter with a certain cross section, the produced lighter dark matter has
a large momentum, namely it is boosted. 
The boosted dark matter behaves as relativistic neutrinos and
may be detected by large neutrino detectors such as
Super-Kamiokande~\cite{Dziomba:2012paz, Ashie:2005ik},
IceCube~\cite{Abbasi:2011eq, Aartsen:2013vca}, and future experiments 
Hyper-Kamiokande~\cite{Kearns:2013lea}, PINGU~\cite{Aartsen:2014oha} and
DUNE~\cite{Acciarri:2015uup} through (in)elastic scattering
with electrons or protons in detectors. 
If the process is inelastic scattering and the decay length of the particle
produced by this scattering is shorter than detector length, another
signal can be expected~\cite{Kim:2016zjx}.
This can be a characteristic signature of multi-component dark matter,
or in other words non-minimal dark sector. 

In this paper, we construct a UV-complete model including two-component dark
matter with a hidden $U(1)_D$ gauge symmetry.
We introduce three new scalar fields and one of these fields has a
milli-charge of the hidden $U(1)_D$ gauge symmetry. 
After the spontaneous symmetry breaking, two hidden particles can be
stable because of the residual $\mathbb{Z}_2\times\mathbb{Z}_2^\prime$
symmetry. 
The lighter component of dark matter has potential to solve small scale
structure problems~\cite{Elbert:2014bma, Tulin:2017ara}, such as
cusp-vs-core problem, too-big-to-fail problem and missing satellite 
problem, with large self-interacting cross section. 
This model is regarded as a variant of the model which has been investigated in the
literature~\cite{Aoki:2016glu}.
The field content is exactly the same, but a difference is that one of the
new particles is milli-charged under the $U(1)_D$ symmetry.
We explore parameter space inducing the signatures of boosted
dark matter in the two-component dark matter model concerning
some relevant constraints such as dark matter relic abundance, direct detection,
cosmic-ray and cosmological observations, perturbativity of couplings,
and we estimate number of multi-Cherenkov ring events originated from boosted dark matter
coming from the Galactic centre which can be detectable at
Hyper-Kamiokande future experiment. 

The paper is organized as follows. In the next Section, the model is
presented in detail.
In Section~\ref{sec:3}, quantitative treatments of boosted
dark matter will be discussed. Section~\ref{sec:4} is devoted to
describe the relevant constraints, and numerical calculations will be
done in Section~\ref{sec:5}. Conclusion and summary is given in
Section~\ref{sec:6}.

%%%%%%%%%%%%%%%
\section{The Model}
\label{sec:2}

\begin{table}[t]
\begin{center}
\caption{Particle content and charge assignment where $Q_\chi$ is an
 arbitrary value consistent with Eq.~(\ref{eq:lag1}) and
 (\ref{eq:lag2}). All the new particles are scalar fields.}
\label{tab:particles}
\begin{tabular}{cccc}\bhline{1pt}
    & $\Sigma$ & $S$ & $\chi$\\\hline
$Q_D$ & $1$ & $-1/2$ & $Q_\chi$\\
Remnant $\mathbb{Z}_2\times\mathbb{Z}_2^\prime$  & $(0,0)$ & $(1,0)$ & (0,1)\\\bhline{1pt}
\end{tabular}
\end{center}
\end{table}

We consider the model extended by a hidden $U(1)_D$ gauge symmetry with
three new scalar fields $\Sigma$, $S$ and $\chi$ as shown in
Tab.~\ref{tab:particles}. 
This model is a variant of the previous model which has
been considered in the literature~\cite{Aoki:2016glu}. 
Unlike the previous case, we take arbitrary hidden $U(1)_D$ charge
for $\chi$ consistent with the following Lagrangian (e.g. $Q_\chi=1/5,2/5,\cdots$) so that the
$\mathbb{Z}_2\times\mathbb{Z}_2^\prime$ symmetry remains after symmetry
breaking.
The kinetic terms of the new scalar fields are given by
\begin{equation}
\mathcal{L}=
\left|D_\mu\Sigma\right|^2+
\left|D_\mu S\right|^2+
\left|D_\mu\chi\right|^2
-\frac{\epsilon}{2}B_{\mu\nu}{Z^{\prime}}^{\mu\nu},
\label{eq:lag1}
\end{equation}
where $\epsilon$ is the kinetic mixing between the Standard Model
$U(1)_Y$ and the hidden $U(1)_D$ symmetries. 
The full scalar potential is written down as
\begin{eqnarray}
\mathcal{V}
\hspace{-0.2cm}&=&\hspace{-0.2cm}
\mu_{\Phi}^2|\Phi|^2+
\mu_{\Sigma}^2|\Sigma|^2+
\mu_{S}^2|S|^2+
\mu_{\chi}^2|\chi|^2
+\frac{\lambda_\Phi}{4}|\Phi|^4
+\frac{\lambda_\Sigma}{4}|\Sigma|^4
+\frac{\lambda_S}{4}|S|^4
+\frac{\lambda_\chi}{4}|\chi|^4\nonumber\\
\hspace{-0.2cm}&&\hspace{-0.2cm}
+\lambda_{\Phi\Sigma}|\Phi|^2|\Sigma|^2
+\lambda_{\Phi S}|\Phi|^2|S|^2
+\lambda_{\Phi\chi}|\Phi|^2|\chi|^2
+\lambda_{\Sigma S}|\Sigma|^2|S|^2
+\lambda_{\Sigma\chi}|\Sigma|^2|\chi|^2
+\lambda_{S\chi}|S|^2|\chi|^2\nonumber\\
\hspace{-0.2cm}&&\hspace{-0.2cm}
 +\left(\frac{\kappa}{2}\Sigma S^2+\mathrm{H.c.}\right).
 \label{eq:lag2}
\end{eqnarray}
In addition to the above scalar potential, the
following terms 
\begin{equation}
\mathcal{V}^\prime=\left(\frac{\mu}{2}S\chi^2+\frac{\lambda}{2}\Sigma
  S{\chi^\dag}^2+\mathrm{H.c.}\right),
\label{eq:add}
\end{equation}
can also be allowed if $Q_\chi=1/4$.\footnote{$Q_\chi=-1/4$ 
also gives the same additional terms with Eq.~(\ref{eq:add}) obtained by exchanging
$\chi\leftrightarrow \chi^\dag$.}
In this case, the model results in the previous model~\cite{Aoki:2016glu}. 
As we will see later, $\chi$ and the CP-odd component of $S$ can be
identified as dark matter candidates simultaneously, and we focus on the dark matter
mass less than 
$\mathcal{O}(10)$ GeV. In this mass range, the magnitude of the
$U(1)_D$ charge 
should be $Q_\chi g_D\lesssim10^{-2}$ in order to be consistent with the observed
relic abundance and direct detection of dark
matter.\footnote{The constraint of dark matter
direct detection can be evaded if $\chi$ splits into CP-even and CP-odd
states since elastic scattering with nuclei does not occur in this case.
Splitting the mass may be achieved by introducing additional scalar
$\Sigma^\prime$ 
coupling with $\chi$ like $\Sigma^\prime\chi^2$ since this term gives a
mass splitting if the field $\Sigma^\prime$ gets a vacuum expectation
value. As an concrete example of charge assignment, one can consider
$Q_\Sigma=1$, $Q_{\Sigma^\prime}=4/5$, $Q_S=-1/2$ and $Q_\chi=-2/5$.} 
Thus the terms in Eq.~(\ref{eq:add}) are
eventually forbidden. 
Hereafter the couplings $\lambda_{\Phi\chi}$ and $\lambda_{\Sigma\chi}$ are
set to be zero for simplicity. 
Practically, these couplings should satisfy
$\lambda_{\Phi\chi}\sin^2\alpha+\lambda_{\Sigma\chi}\cos^2\alpha
\lesssim10^{-4}\left(m_\chi/\mathrm{GeV}\right)$ not to affect
our following analysis (not too deplete relic abundance of $\chi$).

The hidden $U(1)_D$ gauge symmetry is spontaneously broken by the vacuum
expectation value of $\Sigma$ which is parametrized as
$\Sigma=\langle\Sigma\rangle+\sigma/\sqrt{2}$, and 
the hidden gauge boson $Z^\prime$ gets the mass
$m_{Z^\prime}\equiv g_D\langle\Sigma\rangle$. 
Similarly, the Higgs doublet field is written as
$\Phi=(0,\langle\Phi\rangle+\phi^0/\sqrt{2})^T$ in the Unitary gauge. 
The gauge eigenstates $\phi^0$ and $\sigma$ mix with each other via the
coupling $\lambda_{\Phi\Sigma}$, and these gauge eigenstates can be
rewritten as
\begin{equation}
\left(
\begin{array}{c}
\phi^0\\
\sigma
\end{array}
\right)=\left(
\begin{array}{cc}
\cos\alpha & -\sin\alpha\\
\sin\alpha & \cos\alpha
\end{array}
\right)\left(
\begin{array}{c}
h\\
H
\end{array}
\right),\quad
\text{with}\quad
\sin2\alpha=
\frac{4\lambda_{\Phi\Sigma}\langle\Phi\rangle\langle\Sigma\rangle}{m_h^2-m_H^2},
\label{eq:higgs_diag}
\end{equation}
where the mass eigenstates $h$ and $H$ correspond to the SM-like Higgs boson
with $m_h=125~\mathrm{GeV}$ and the second extra Higgs boson, respectively. 
With the orthogonal matrix given in Eq.~(\ref{eq:higgs_diag}), one can
rewrite the quartic couplings $\lambda_\Phi$ and $\lambda_\Sigma$ as 
\begin{eqnarray}
\lambda_{\Phi}
\hspace{-0.2cm}&=&\hspace{-0.2cm}
\frac{1}{\langle\Phi\rangle^2}\left(
m_h^2\cos^2\alpha + m_H^2\sin^2\alpha
\right),\\
\lambda_{\Sigma}
\hspace{-0.2cm}&=&\hspace{-0.2cm}
\frac{1}{\langle\Sigma\rangle^2}\left(
m_h^2\sin^2\alpha + m_H^2\cos^2\alpha
\right).
\end{eqnarray}
The complex scalar field $S=\left(s+ia\right)/\sqrt{2}$ splits into the
CP-even and CP-odd mass eigenstates $s$ and $a$ whose masses are given by 
\begin{eqnarray}
m_s^2
\hspace{-0.2cm}&=&\hspace{-0.2cm}
\mu_S^2+\lambda_{\Sigma
S}\langle\Sigma\rangle^2+\kappa\langle\Sigma\rangle,
\label{eq:smass}\\
m_a^2
\hspace{-0.2cm}&=&\hspace{-0.2cm}
\mu_S^2+\lambda_{\Sigma
S}\langle\Sigma\rangle^2-\kappa\langle\Sigma\rangle.
\label{eq:amass}
\end{eqnarray}
Thus the squared mass difference is given by
$m_s^2-m_a^2=2\kappa\langle\Sigma\rangle$. 
Since we can take $\kappa>0$ without loss of generality, this means $m_s>m_a$.

\begin{figure}[t]
\begin{center}
\includegraphics[scale=0.7]{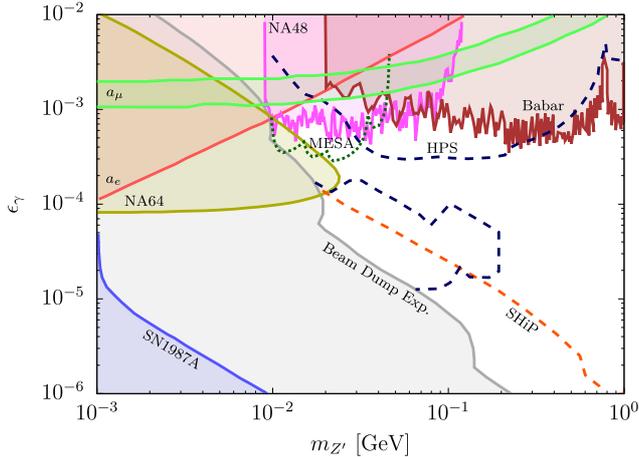}
\caption{Current limits on the kinetic mixing
 $\epsilon_\gamma$ including beam dump
 experiments~\cite{Essig:2013lka},
 SN1987A~\cite{Kazanas:2014mca}, NA48/2~\cite{Batley:2015lha},
 BaBar~\cite{Lees:2014xha} and NA64~\cite{Banerjee:2018vgk}. The dotted
 lines correspond to the future experiments: HPS~\cite{Essig:2013lka},
 MESA~\cite{Beranek:2013yqa} and SHiP~\cite{Alekhin:2015byh}.} 
\label{fig:eps_mzp}
\end{center}
\end{figure}

The gauge kinetic mixing $\epsilon_\gamma\equiv\epsilon\cos\theta_W$ is
experimentally 
constrained by $Z^\prime\to e^+e^-$ search and the current exclusion
limits on the kinetic mixing 
$\epsilon_\gamma$ are summarized in Fig.~\ref{fig:eps_mzp} including beam dump
experiments~\cite{Essig:2013lka},
SN1987A~\cite{Kazanas:2014mca}, NA48/2~\cite{Batley:2015lha},
BaBar~\cite{Lees:2014xha} and NA64~\cite{Banerjee:2018vgk}.\footnote{In our
case, since the main decay mode of $Z^\prime$ would be $Z^\prime\to
sa\to e^+e^-aa$, the exclusion limits in Fig.~\ref{fig:eps_mzp}
cannot directly be applied. However one can take these limits as 
conservative limits.}
The light red region is excluded by the measurement of the electron
anomalous magnetic moment. 
The green region can explain the discrepancy of muon anomalous magnetic
moment with the Standard Model, but all the region is excluded by the
other limits. 
According to the plot, the upper limit on the kinetic mixing is roughly given by
$\epsilon_\gamma\lesssim10^{-3}$ for $m_{Z^\prime}\gtrsim10~\mathrm{MeV}$.
The Higgs mixing angle $\sin\alpha$ is also constrained by some experiments,
and the strong bound $\sin\alpha\lesssim0.01$ is given by B-meson decays
when the second Higgs mass is $m_{H}\lesssim5~\mathrm{GeV}$~\cite{Falkowski:2015iwa}.
Hereafter we fix the kinetic mixing
parameter to be $\epsilon_\gamma=5\times10^{-4}$ which can be explored by
the HPS future experiment~\cite{Essig:2013lka}, and the Higgs mixing
angle is also fixed to be
$\sin\alpha=10^{-3}$.

The hidden $U(1)_D$ gauge symmetry is broken by the vacuum expectation value
of $\Sigma$. After the symmetry breaking, a pair of two particles ($s$,
$\chi$) or ($a$, $\chi$) is stabilized by the residual
$\mathbb{Z}_2\times\mathbb{Z}_2^\prime$ symmetry.
Since $m_s>m_a$ because of the positive $\kappa$ coupling that we assumed, a pair
of $a$ and $\chi$ can be stable dark matter particles, namely
two-component dark matter. 
Furthermore, if the annihilation cross section for $\chi^\dag\chi\to
aa~(m_a<m_\chi)$ is large enough, the lighter dark matter $a$ is boosted
at a region of high dark matter density (the Galactic centre for example).
Such boosted dark matter can be detectable by large neutrino detectors
such as Super-Kamiokande, IceCube, Hyper-Kamiokande, PINGU and DUNE. 
In the rest of the paper, we focus on the following mass interval and
mass hierarchy
\begin{equation}
10~\mathrm{MeV}\lesssim m_{a}\lesssim m_{H},~m_{s}\lesssim
 m_{Z^\prime}\lesssim m_{\chi}\lesssim10~\mathrm{GeV},
\label{eq:mass_hierarchy}
\end{equation}
in order to maximize possible signature of the boosted dark matter and
obtain a large self-interacting cross section of the lighter dark matter $a$
to improve the small scale structure problems such as
cusp-vs-core problem, too-big-to-fail problem and missing satellite
problem as we will
discuss later.

The fraction of the heavier component of dark matter $\chi$ should be
large to enhance the signals of the boosted dark matter via the
annihilation $\chi^\dag\chi\to aa$ 
while the fraction of the lighter component of dark matter $a$ should
also be large to improve the small scale structure problems with a large
self-interacting cross section. 
Therefore it is interesting to investigate the case that both components of dark
matter are comparable. The concrete calculation of the relic abundance
will be performed in Section~\ref{sec:4}, and we will concentrate on the
parameter space such that each relic abundance $\Omega_\chi h^2$ and 
$\Omega_a h^2$ is in the range $40\%\sim60\%$ of the observed total
abundance which is taken as $3\sigma$ range of the PLANCK Collaboration
$\Omega_{\chi}h^2+\Omega_ah^2=0.1197\pm0.0022$~\cite{Ade:2015xua}.

\section{Signatures of the Boosted Dark Matter}
\label{sec:3}
In the non-relativistic limit, the thermally averaged annihilation cross
section for $\chi^\dag\chi\to aa$ is simply given by 
\begin{equation}
\langle\sigma{v}\rangle_{\chi^\dag\chi\to aa}
=\frac{\lambda_{S\chi}^2}{64\pi m_\chi^2}
\sqrt{1-\frac{m_a^2}{m_\chi^2}}.
\label{eq:xxaa}
\end{equation}
Note again the assumption of the scalar couplings
 $\lambda_{S\chi}\gg\lambda_{\Phi\chi},\lambda_{\Sigma\chi}$ as our setup. 
The boosted dark matter $a$ behaves as relativistic neutrinos and can be
detectable with large size neutrino detectors via scattering with
the electrons or protons in the detectors.
In particular, the Super-Kamiokande Collaboration has recently
investigated on the constraint of the boosted dark matter~\cite{Kachulis:2017nci}.
Hereafter we focus on scattering with electrons in water Cherenkov
detectors since the expected number of events for proton scattering 
is rather smaller than that for electrons due to kinematics. 
%%%

\begin{figure}[t]
\begin{center}
\includegraphics[scale=1.2]{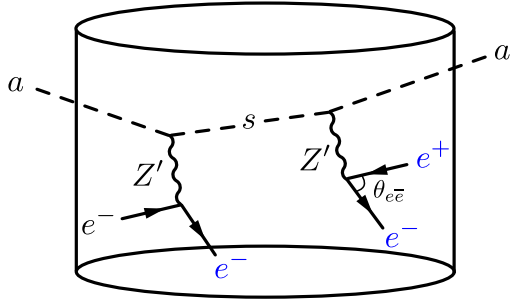}
\caption{Two or three distinctive signals (electrons and positron) of
 the boosted dark matter $a$ at 
 Super-Kamiokande and Hyper-Kamiokande. The blue electron and position and its angle
 $\theta_{e\overline{e}}$ can be detected.}
\label{fig:boosted}
\end{center}
\end{figure}

In principle, the CP-even state $s$ is also produced by
the annihilation $\chi^\dag\chi\to ss$ with a similar production cross
section of Eq.~(\ref{eq:xxaa}), and then the CP-even state $s$
immediately decays into $a$. 
Thus the total energy distribution of the boosted dark matter $a$ would be
described by sum of those two contributions. 
However since energy of $a$ for the latter case is rather small and
is likely to be smaller than experimental energy threshold, we neglect
the contribution.\footnote{Taking into account
the contribution of the $s$ decay would be necessary if a 
more detailed phenomenological feature of the model is required to
discriminate from the other models.} 

The schematic picture in the detector is shown in Fig.~\ref{fig:boosted}.
First, a boosted dark matter $a$ comes in the detector, and the
excited state $s$ is produced by the inelastic scattering $ae^-\to
se^-$.
The produced $s$ subsequently decays
through the mode $s\to ae^+e^-$.\footnote{Depending on the masses
of $a$ and $s$, another decay mode $s\to a\mu^+\mu^-$ may also be possible.} 
If the decay length defined by $L_D\equiv\beta_s\gamma_s\tau_s$ is shorter
than the detector length, multi-Cherenkov ring events can be observed where
$\gamma_s= E_s/m_s$ is the boost factor,
$\beta_s=|\bm{p}_s|/E_s$ is the velocity and
$\tau_s\equiv\Gamma_{s\to ae^+e^-}^{-1}$ is the lifetime of the
excited state $s$.
The detector length (diameter of detector) is $39.3~\mathrm{m}$ for
Super-Kamiokande and $74~\mathrm{m}$ for Hyper-Kamiokande. 
%%%
Furthermore if the angle between $e^+$ and $e^-$ generated by the $s$ decay
($\theta_{e\overline{e}}$ shown in Fig.~\ref{fig:boosted}) is larger than
experimental threshold 
($\theta_{e\overline{e}}^\mathrm{exp}\gtrsim3^\circ$) at Super-Kamiokande and
Hyper-Kamiokande, a 3-Cherenkov ring event is expected to be observed. Otherwise,
a 2-Cherenkov ring event is expected. 
These can be characteristic signals of non-minimal dark sector~\cite{Kim:2016zjx}.
Note that although we do not consider scattering with a proton in this paper, an
advantage of proton scattering is that the electron and positron 
induced by the decay $s\to ae^+e^-$ may be easily distinguishable since
the angle between those 
particles are likely to be larger than the case of electron scattering due to
rather small boost factor $\gamma_s$ for proton
scattering~\cite{Agashe:2014yua, Kim:2016zjx}.

We can roughly estimate the total number of multi-ring events which has
been observed by Super-Kamiokande.
From the literature~\cite{Nishino:2009aa}, one can find that $3036$
$e$-like multi-ring events have totally observed with $140.9$ kiloton-year exposure. 
This is translated into $485$ events per year at Super-Kamiokande. 
Among the total number of the events, only the events relevant to our
multi-ring signals should be extracted by taking into account the
direction, position and energy of the events in order to be compared with
the prediction in the model. 
However since such detailed analysis of multi-Cherenkov ring events coming from
Galactic centre at Super-Kamiokande has not been done,
there is no substantial experimental bound for the boosted dark matter signals of
multi-component dark matter.\footnote{We leave a systematic background analysis to the future work.} 

The differential cross section for the inelastic scattering $ae^-\to
se^-$ mediated by $Z^\prime$ is computed as
\begin{equation}
\frac{d\sigma_\mathrm{inel}}{dE_e}=
\frac{m_e\overline{|\mathcal{M}_\mathrm{inel}|^2}}
{8\pi\lambda(s,m_{a}^2,m_e^2)},
\label{eq:dinel}
\end{equation}
where $s$ is the Manderstam variable related with the other
variables $t$ and $u$ as $s+t+u=m_a^2+m_s^2+2m_e^2$,
$\lambda(x,y,z)$ is the 
kinematical function defined by
$\lambda(x,y,z)=x^2+y^2+z^2-2xy-2yz-2zx$, $E_e$ is the energy of the scattered
electron and $\overline{|\mathcal{M}_\mathrm{inel}|^2}$ is the squared
amplitude spin averaged over initial state and summed over final state
which is given by 
\begin{equation}
\overline{|\mathcal{M}_\mathrm{inel}|^2}
=\frac{g_D^2\epsilon_\gamma^2\alpha_\mathrm{em}\pi}{(t-m_{Z^\prime}^2)^2}
\Bigl[
(s-u)^2-t^2+2t(m_{s}^2+m_{a}^2)-(m_{s}^2-m_{a}^2)^2
\Bigr].
\end{equation}
The Manderstam variable $t$ is correlated with the scattered electron energy
as $t=2m_e(m_e-E_e)$. 
The total cross section is obtained by integrating Eq.~(\ref{eq:dinel})
with respect to $E_e$.
For a given energy $s$, the minimal and maximal $E_e$ are given by 
\begin{eqnarray}
E_e^\mathrm{max}
\hspace{-0.2cm}&=&\hspace{-0.2cm}
\frac{\left(s-m_a^2+m_e^2\right)\left(s-m_s^2+m_e^2\right)}{4m_es}
+\frac{\sqrt{\lambda\left(s,m_a^2,m_e^2\right)}\sqrt{\lambda\left(s,m_{s}^2,m_e^2\right)}}
{4m_es},\\
E_e^\mathrm{min}
\hspace{-0.2cm}&=&\hspace{-0.2cm}
\frac{\left(s-m_a^2+m_e^2\right)\left(s-m_s^2+m_e^2\right)}{4m_es}
-\frac{\sqrt{\lambda\left(s,m_a^2,m_e^2\right)}\sqrt{\lambda\left(s,m_{s}^2,m_e^2\right)}}
{4m_es}.
\end{eqnarray}
For comparison with Super-Kamiokande and Hyper-Kamiokande experiments, the
experimental energy thresholds should be taken into account. 
The actual energy threshold of these experiments is
$E_e^\mathrm{exp}\gtrsim0.01~\mathrm{GeV}$.
However since angular resolution is not so good for lower
energy, we take a conservative lower energy threshold
$E_e^\mathrm{exp}>0.1~\mathrm{GeV}$~\cite{Agashe:2014yua}.

The decay width for $s\to af\overline{f}$ mediated by off-shell
$Z^\prime$ is calculated as 
\begin{equation}
\Gamma_{s\to af\overline{f}}=
\frac{g_D^2\alpha_\mathrm{em}\epsilon_\gamma^2}{192\pi^2 m_s^3}
\int_{4m_f^2}^{(m_s-m_a)^2}
\frac{\sqrt{(q^2-m_s^2-m_a^2)^2-4m_s^2m_a^2}^3}
{(q^2-m_{Z^\prime}^2)^2+m_{Z^\prime}^2\Gamma_{Z^\prime}^2}
\sqrt{1-\frac{4m_f^2}{q^2}}\left(1+\frac{2m_f^2}{q^2}\right)
dq^2,
\end{equation}
where $q^2\equiv(p_s-p_a)^2$ is transfer momentum
of $Z^\prime$, and $\Gamma_{Z^\prime}$ is the total decay width of
$Z^\prime$ which is given by 
\begin{eqnarray}
\Gamma_{Z^\prime}
\hspace{-0.2cm}&=&\hspace{-0.2cm}
\sum_{f}\frac{\alpha_\mathrm{em}\epsilon_\gamma^2m_{Z^\prime}}{3}
\left(1+2\frac{m_f^2}{m_{Z^\prime}^2}\right)
\sqrt{\lambda\left(1,\frac{m_f^2}{m_{Z^\prime}^2},\frac{m_f^2}{m_{Z^\prime}^2}\right)}
\nonumber\\
\hspace{-0.2cm}&&\hspace{-0.2cm}
+\frac{g_D^2m_{Z^\prime}}{192\pi}
\left[
1-2\left(\frac{m_s^2}{m_{Z^\prime}^2}+\frac{m_a^2}{m_{Z^\prime}^2}\right)+\left(\frac{m_s^2}{m_{Z^\prime}^2}-\frac{m_a^2}{m_{Z^\prime}^2}\right)^2
\right]
\sqrt{\lambda\left(1,\frac{m_s^2}{m_{Z^\prime}^2},\frac{m_a^2}{m_{Z^\prime}^2}\right)},
\label{eq:zp_decay}
\end{eqnarray}
with the kinematical function $\lambda(x,y,z)$ defined above.
The first term in Eq.~(\ref{eq:zp_decay}) corresponds to the decay mode
$Z^\prime\to f\overline{f}$ 
and the second term is $Z^\prime\to sa$. 
Since the kinetic mixing $\epsilon_\gamma$ is small, the decay width for
$Z^\prime\to f\overline{f}$ is suppressed and the contribution of  
$Z^\prime\to sa$ becomes dominant if it is kinematically allowed.

The energy of the incoming boosted dark matter $a$ denoted by $E_a$
should be as large as
\begin{equation}
E_a>\frac{m_s^2-m_a^2+2m_sm_e}{2m_e},
\label{eq:inel}
\end{equation}
so that the inelastic scattering $ae^-\to se^-$ can be kinematically
accessible. 
The energy of the boosted dark matter is given by the mass of the heavier
component dark matter ($E_a\approx m_\chi$) since the dark matter
particles are non-relativistic. 
Eq.~(\ref{eq:inel}) implies that a larger energy $E_a$
is kinematically required for scattering with electrons compared to that
with protons 
because of the target particle mass in the denominator of Eq.~(\ref{eq:inel}) ($m_e\ll m_p$).
On the other hand, if the energy of the initial state $a$ is
too large, the decay length of the excited state $s$ 
tends to be too long to decay inside the detectors. 
Therefore one can expect that relevant parameter space for
multi-Cherenkov ring events is bounded from both below and above.

It is expected that the annihilation $\chi^\dag\chi\to aa$ occurs at the
Galactic centre, and the dark matter $a$ is produced in all the directions.
We take the dark matter flux only within $10^\circ$ cone around the
Galactic centre. In this case, the flux of the boosted dark matter $a$ is
estimated by~\cite{Agashe:2014yua}
\begin{equation}
\Phi_{a}^{10^\circ}=9.9\times10^{-8}~\mathrm{cm^{-2}s^{-1}}
\left(\frac{\langle\sigma{v}\rangle_{\chi^\dag\chi\to aa}}
{5\times10^{-26}~\mathrm{cm^3/s}}\right)
\left(\frac{20~\mathrm{GeV}}{m_\chi}\right)^2
\left(\frac{\Omega_\chi}{\Omega_a+\Omega_\chi}\right)^2,
\end{equation}
where the fraction of total relic abundance
$\Omega_\chi/(\Omega_a+\Omega_\chi)$ is multiplied, and the
Navarro-Frenk-White (NFW) dark
matter profile is assumed~\cite{Navarro:1995iw}. 
The energy distribution of the scattered electron in the detector is
given by~\cite{Agashe:2014yua}
\begin{equation}
 E_e\frac{dN_\mathrm{inel}}{dE_e}=\delta{t} N_\mathrm{target}\Phi_{a}^{10^\circ}E_e
  \frac{d\sigma_\mathrm{inel}}{dE_e},
\end{equation}
where $\delta{t}$ is the exposure time, $N_\mathrm{target}$ is the number of
target electrons in the detector which can be estimated as
$N_\mathrm{target}=7.49\times10^{33}$ for Super-Kamiokande and
$N_\mathrm{target}=1.25\times10^{35}$ for Hyper-Kamiokande from the
fiducial volume $0.0224$ megaton and
$0.187\times2=0.374$ megaton (taking into
account two 0.187 megaton tanks)~\cite{Abe:2018uyc}, respectively.
Integrating over the electron energy, the total expected number of
events for the inelastic scattering is obtained. 

Furthermore, we define the number of the (multi-)Cherenkov ring events in a
specific region of interest as
\begin{equation}
 N_\mathrm{signal}=\delta{t}N_\mathrm{target}\Phi_a^{10^\circ}
  \!\!\int_{E_e^\mathrm{min}}^{E_e^\mathrm{max}}\!\!dE_e
  \frac{d\sigma_\mathrm{inel}}{dE_e}
  \mathrm{Br}_{s\to ae\overline{e}}
  \!\!\int_{{E_e^\prime}^\mathrm{min}}^{{E_e^\prime}^\mathrm{max}}\!\!dE_e^\prime
  \!\!\int_{{E_{\overline{e}}}^\mathrm{min}}^{{E_{\overline{e}}}^\mathrm{max}}\!\!dE_{\overline{e}}
  \!\!\int_{\theta_{e\overline{e}}^\mathrm{min}}^{\theta_{e\overline{e}}^\mathrm{max}}\!\!d\theta_{e\overline{e}}
  \frac{d^3N_{s}}{dE_e^\prime dE_{\overline{e}}d\theta_{e\overline{e}}},
\end{equation}
where $\mathrm{Br}_{s\to ae\overline{e}}$ is the branching ratio of the decay
$s\to ae^+e^-$ which is $\mathrm{Br}_{s\to ae\overline{e}}=1$ in our
case, $E_e^\prime$ and $E_{\overline{e}}$ are the energy of the electron
and positron produced by the decay of the boosted CP-even state $s$, and
$d^3N_{s}/dE_e^\prime
dE_{\overline{e}}d\theta_{e\overline{e}}$ is the energy and angular 
distribution of the electron and positron normalized to $N_{s}=1$. 
%%%
In order to obtain the distribution
$d^3N_{s}/dE_e^\prime dE_{\overline{e}}d\theta_{e\overline{e}}$ for the boosted state
$s$, first we generate events of the $s$ decay at rest frame with
CalcHEP~\cite{Pukhov:1999gg, Pukhov:2004ca}, then each event is boosted
by the Lorentz transformation.

\section{The Constraints}
\label{sec:4}
\subsection{Relic Abundance of Dark Matter}
In general, one has to solve the coupled Boltzmann equation to calculate
the relic abundance of the two-component dark matter. 
However since the masses of the dark matter particles are 
hierarchical ($m_a\ll m_\chi$), the coupled Boltzmann equation can be
simplified as follows 
\begin{eqnarray}
\frac{dn_\chi}{dt}+3Hn_\chi
\hspace{-0.2cm}&=&\hspace{-0.2cm}
-\frac{\langle\sigma{v}\rangle_{\chi^{\dag}\chi}}{2}\left(n_\chi^2-{n_\chi^\mathrm{eq}}^2\right),
\label{eq:boltzmann1}\\
\frac{dn_a}{dt}+3Hn_a
\hspace{-0.2cm}&=&\hspace{-0.2cm}
-\langle\sigma{v}\rangle_{aa}\left(n_a^2-{n_a^\mathrm{eq}}^2\right),
\label{eq:boltzmann2}
\end{eqnarray}
where $n_\chi$ is the total number density of $\chi$ and
$\chi^\dag$,\footnote{Note that anti-dark matter particle $\chi^\dag$ is
different degree of freedom from dark matter $\chi$. The number
densities of $\chi$ and $\chi^\dag$ can be regarded as exactly same
assuming CP invariance. Because of this reason, the factor $1/2$ appears
in Eq.~(\ref{eq:boltzmann1}).}
$n_a$ is the
dark matter number density of $a$, 
$n_\chi^\mathrm{eq}$ and $n_a^\mathrm{eq}$ are the number densities in
thermal equilibrium, $H$ is the Hubble
parameter, and $\langle\sigma{v}\rangle_{\chi^{\dag}\chi}$ and
$\langle\sigma{v}\rangle_{aa}$ are the total annihilation cross
sections for each dark matter. 
These equations can be independently solved, and the total relic
abundance should satisfy $\Omega_ah^2+\Omega_\chi h^2\approx0.12$
observed by the PLANCK Collaboration~\cite{Ade:2015xua}. 

\begin{figure}[t]
\begin{center}
\includegraphics[scale=0.7]{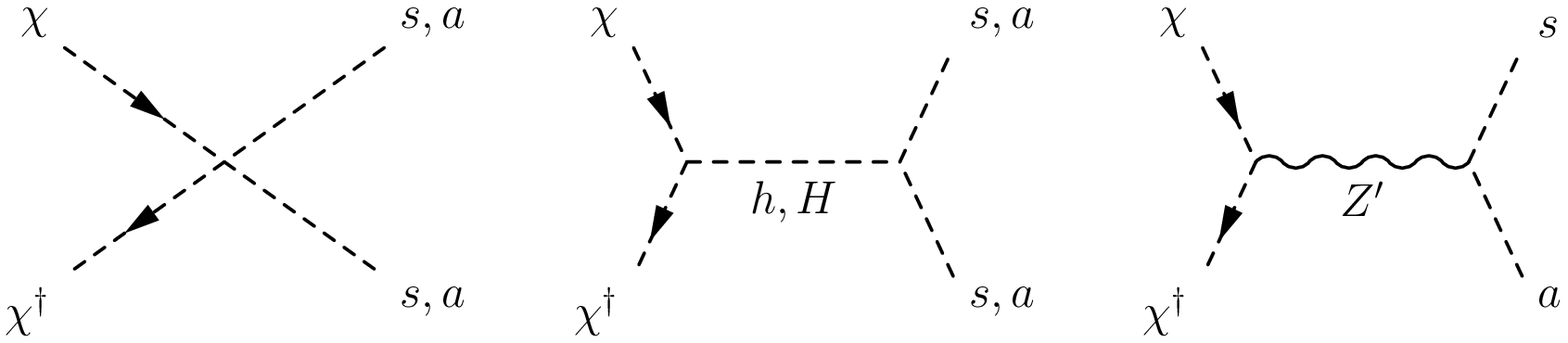}
\includegraphics[scale=0.7]{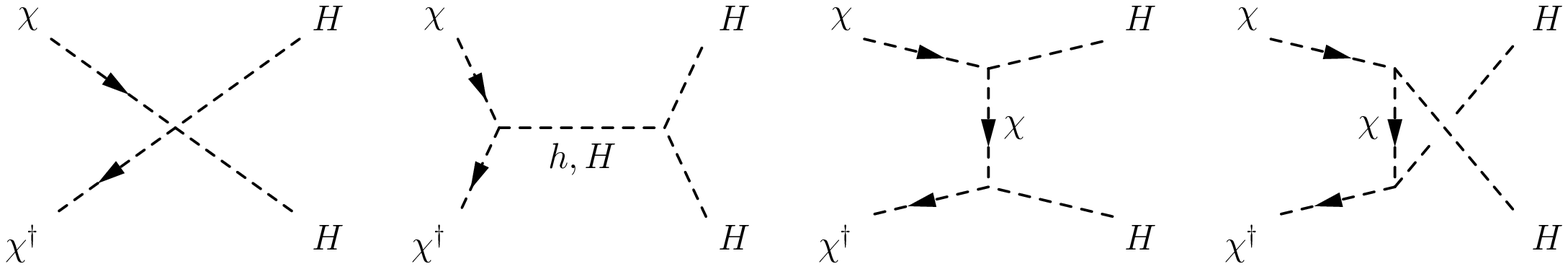}
\includegraphics[scale=0.7]{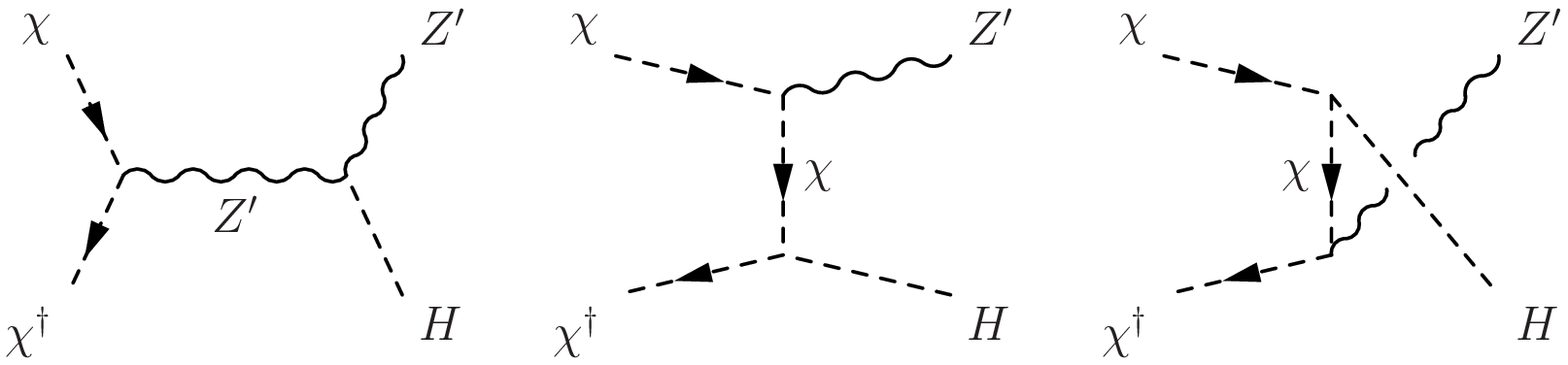}
\includegraphics[scale=0.7]{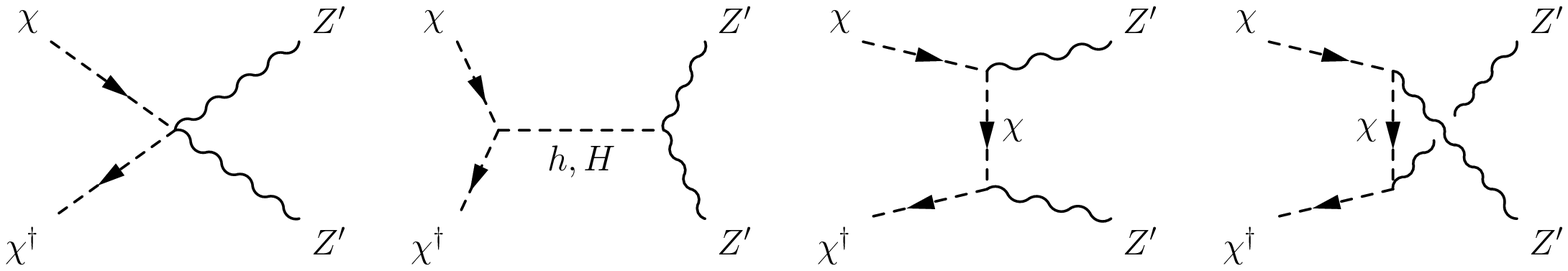}
\includegraphics[scale=0.7]{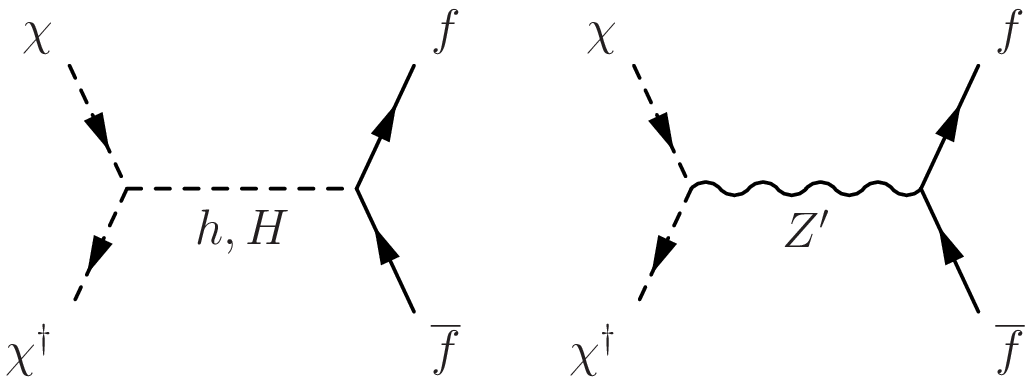}
\caption{Complete diagrams for $\chi$ annihilations.}
\label{fig:chi_ann}
\end{center} 
\end{figure}

The possible annihilation channels for $\chi$ dark matter are
$\chi^\dag\chi\to ss$, $aa$, $sa$, $HH$, $HZ^\prime$, $Z^\prime
Z^\prime$, $f\overline{f}$ where $f$ is a Standard Model fermion and the 
relevant diagrams are shown in Fig.~\ref{fig:chi_ann}. 
Since we have assumed that $\lambda_{\Phi\chi}$ and
$\lambda_{\Sigma\chi}$ 
are sub-dominant, some diagrams are negligible. 
In particular, the annihilation channel $\chi^\dag\chi\to HH$ (the
second line of Fig.~\ref{fig:chi_ann}) completely disappears in the case
of $\lambda_{\Phi\chi}=\lambda_{\Sigma\chi}=0$. 
In fact, this is required in
order to be consistent with cosmic-ray and Cosmic Microwave Background (CMB) observations as we will
see later. This is because if the annihilation cross section for the
channel $\chi^\dag\chi\to HH$ is large enough, the produced second Higgs
boson $H$ subsequently decays into $e^+e^-$. 
The other two channels $\chi^\dag\chi\to Z^\prime Z^\prime$, $HZ^\prime$
should also be suppressed due to the same reason while the main decay
channel of $Z^\prime$ is different ($Z^\prime\to sa\to e^+e^-aa$).
As a result, the coupling $|g_DQ_\chi|$ is strongly constrained.
In addition, dark matter direct detection also gives strong constraint on $|g_DQ_\chi|$. 
Thus one can see that the main annihilation channel should eventually be
$\chi^\dag\chi\to ss$, $aa$ in
the mass range given by Eq.~(\ref{eq:mass_hierarchy}). 
We use the public code micrOMEGAs to compute the relic abundance of
$\chi$~\cite{Belanger:2014vza}.

\begin{figure}[t]
\begin{center}
\includegraphics[scale=0.7]{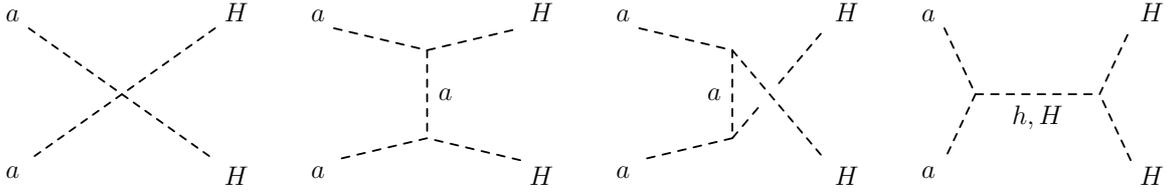}
\caption{Complete diagrams for $a$ annihilations. This annihilation
 is a forbidden channel due to $m_a<m_H$ but can occur with certain
 momentum in the early universe.}
\label{fig:a_ann}
\end{center} 
\end{figure}

For the lighter dark matter $a$, the relic abundance can be determined
by the forbidden 
channel $aa\to HH$ where $m_a<m_H$.\footnote{The 3-to-2 process $aaa\to aH$
may give an impact on the computation of the relic
abundance of $a$~\cite{Dey:2016qgf, Cline:2017tka}.} 
The relevant diagrams are shown in Fig.~\ref{fig:a_ann}. 
In non-relativistic limit, 
the annihilation cross section for the forbidden channel is correlated
with the inverse process $HH\to aa$~\cite{DAgnolo:2015ujb}, and is given by
\begin{equation}
\langle\sigma{v}\rangle_{aa\to HH}=
\frac{m_H}{32\pi m_a^3}
\left|
\lambda_{HHaa}-\frac{2\mu_{Haa}^2}{m_H^2}
+\frac{\mu_{Haa}\mu_{HHH}}{3m_H^2+im_H\Gamma_H}
\right|^2
\sqrt{1-\frac{m_a^2}{m_H^2}}e^{-2(m_H-m_a)/T},
\label{eq:forbidden}
\end{equation}
where $T$ is temperature of the universe
and the couplings $\lambda_{HHaa}$, $\mu_{Haa}$ $\mu_{HHH}$ are given by
\begin{eqnarray}
\lambda_{HHaa}
\hspace{-0.2cm}&=&\hspace{-0.2cm}
\lambda_{\Phi S}\sin^2\alpha+\lambda_{\Sigma S}\cos^2\alpha,\\
%%%%%
\mu_{Haa}
\hspace{-0.2cm}&=&\hspace{-0.2cm}
-\sqrt{2}\lambda_{\Phi S}\langle\Phi\rangle\sin\alpha
+\sqrt{2}\lambda_{\Sigma S}\langle\Sigma\rangle\cos\alpha
-\frac{\kappa}{\sqrt{2}}\cos\alpha,\label{eq:Haa}\\
%%%%%
\mu_{HHH}
\hspace{-0.2cm}&=&\hspace{-0.2cm}
\frac{3}{\sqrt{2}}\Bigl(
-\lambda_\Phi\langle\Phi\rangle\sin^3\alpha
+\lambda_\Sigma\langle\Sigma\rangle\cos^3\alpha
\Bigr)\nonumber\\
&&
+\frac{6}{\sqrt{2}}\lambda_{\Phi\Sigma}\sin\alpha\cos\alpha
\Bigl(
-\langle\Phi\rangle\cos\alpha
+\langle\Sigma\rangle\sin\alpha
\Bigr).
\end{eqnarray}
For the $H$ decay, the possible decay channel at tree level is $H\to
f\overline{f}$ and $H\to aa$ if $m_H>2m_a$ in our setup and the total
decay width is given by 
\begin{equation}
\Gamma_H=\sum_{f}\frac{y_f^2m_H\sin^2\alpha}{16\pi}\left(1-4\frac{m_f^2}{m_H^2}\right)^{3/2}
+\frac{\mu_{Haa}^2}{32\pi m_H}\sqrt{1-\frac{4m_a^2}{m_H^2}},
\label{eq:H_decay}
\end{equation}
where $y_f$ is the Yukawa coupling. 
The former term in Eq.~(\ref{eq:H_decay}) corresponds to the decay
channel of $H\to f\overline{f}$ and the latter term is $H\to aa$. 
One can find that the cross section in Eq.~(\ref{eq:forbidden}) is
exponentially suppressed if $(m_H-m_a)/T\gg1$.
The Boltzmann equation for the lighter dark matter $a$
given by Eq.~(\ref{eq:boltzmann2}) is numerically solved and the relic
abundance can be computed. 

\subsection{Direct Detection of Dark Matter}
\begin{figure}[t]
\begin{center}
\includegraphics[scale=0.65]{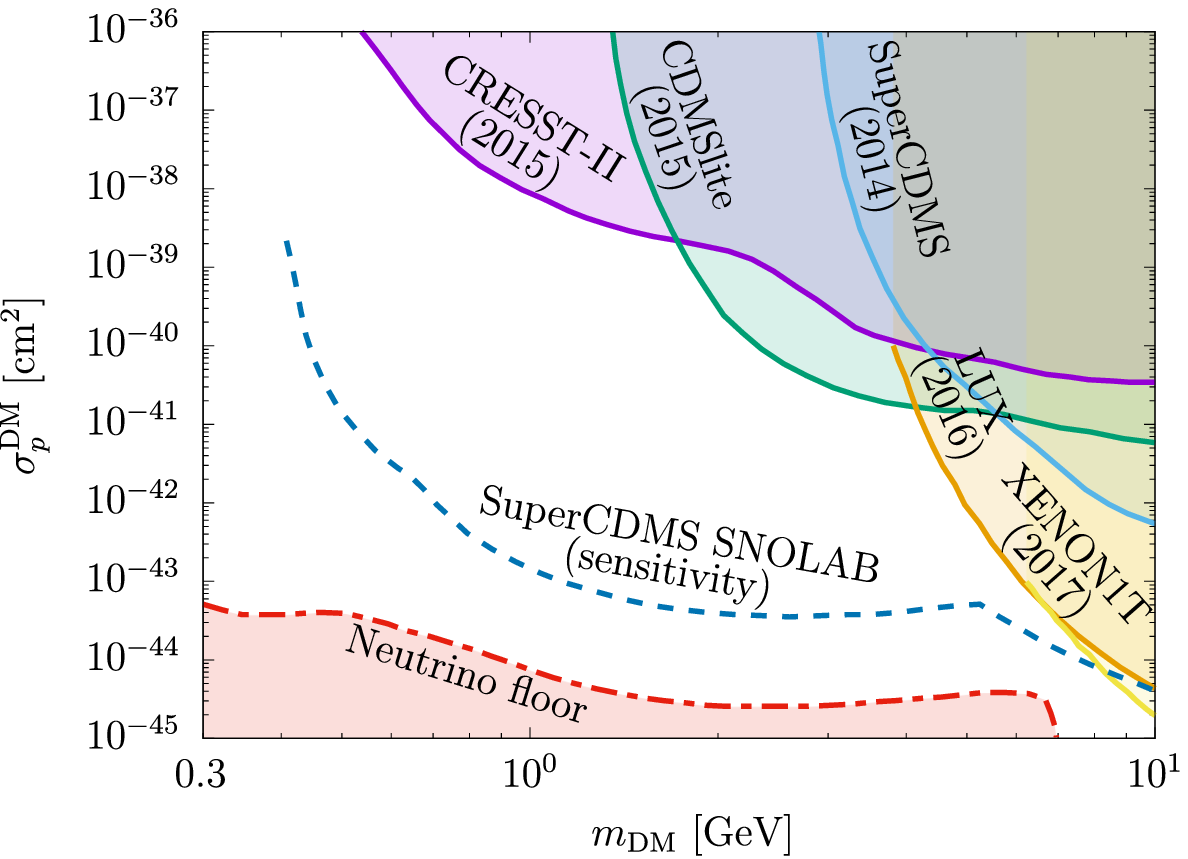}
\includegraphics[scale=0.65]{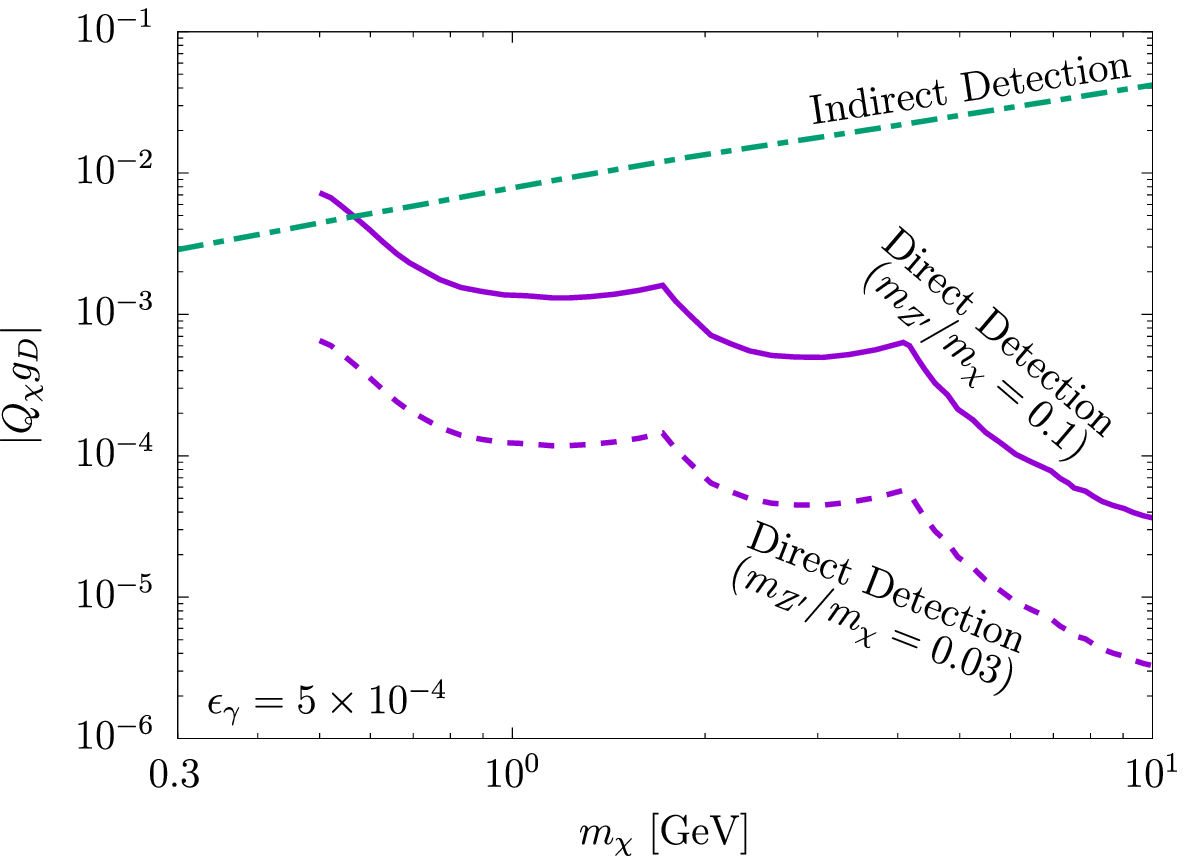}
\caption{(Left): Current direct detection bounds
 (CRESST-II~\cite{Angloher:2015ewa}, CDMSlite~\cite{Agnese:2015nto},
SuperCDMS~\cite{Agnese:2014aze}, LUX~\cite{Akerib:2015rjg},
XENON1T~\cite{Aprile:2017iyp}) and future sensitivities (SuperCDMS SNOLAB~\cite{Agnese:2016cpb}).
(Right): Bounds for the hidden $U(1)_D$ charge obtained from direct
 detection and indirect detection of the heavier component of dark
 matter for $f_\chi=0.5$. The indirect detection bound is insensitive with the mass
 hierarchy $m_{Z^\prime}/m_\chi$ unless $\chi$ and $Z^\prime$ are highly
 degenerate.}
\label{fig:dd_exp}
\end{center}
\end{figure}

The current direct detection bounds and future sensitivities for dark
matter mass range less than 10 GeV are summarized in the left plot of
Fig.~\ref{fig:dd_exp}. 
The purple, green, blue, orange and yellow regions are already excluded by the
experiments CRESST-II~\cite{Angloher:2015ewa}, CDMSlite~\cite{Agnese:2015nto},
SuperCDMS~\cite{Agnese:2014aze}, LUX~\cite{Akerib:2015rjg},
XENON1T~\cite{Aprile:2017iyp}, respectively. 
The blue dotted line represents the future sensitivity of
SuperCDMS SNOLAB~\cite{Agnese:2016cpb}. 
In the red region in the bottom, dark matter scattering cannot be
distinguished from the elastic scattering with neutrinos (so-called
neutrino floor). 

The elastic scattering between the heavier dark matter $\chi$ and a
proton is induced by the $Z^\prime$ boson via the gauge kinetic mixing
$\epsilon_\gamma$, and the spin independent cross section is computed as
\begin{equation}
\sigma_p^\chi=
\frac{4Q_\chi^2g_D^2\alpha_\mathrm{em}\epsilon_\gamma^2m_p^2m_\chi^2}{m_{Z^\prime}^4(m_p+m_\chi)^2},
\label{eq:dd1}
\end{equation}
where $m_p=938~\mathrm{MeV}$ is the proton mass. 
We define an effective scattering cross section
${\sigma_p^\chi}^\mathrm{eff}\equiv\sigma_p^\chi f_\chi$ to compare
with the experimental bounds where
$f_\chi\equiv\Omega_\chi/\left(\Omega_a+\Omega_\chi\right)$. 
This cross section is enhanced by the light mediator mass $m_{Z^\prime}$
when $m_{Z^\prime}\ll m_\chi$.
Although the experimental direct detection bound for
$m_\chi\lesssim10~\mathrm{GeV}$ is rather weaker than that for the case of
$m_\chi\gtrsim10~\mathrm{GeV}$, the constraint is strong enough since we are
interested in the region of small mediator mass ($m_{Z^\prime}\ll m_\chi$). 
With Eq.~(\ref{eq:dd1}), the current upper bound on the elastic cross
section can be translated into an upper bound on $|Q_\chi g_D|$ as shown
in the right panel (purple lines) in Fig.~\ref{fig:dd_exp} for $f_\chi=0.5$.
The upper bound depends on mass hierarchy $m_{Z^\prime}/m_\chi$. 

The dark matter $\chi$ can also scatter electron off with the same
formula of the elastic cross section in Eq.~(\ref{eq:dd1}) where $m_p$
is replaced to $m_e$. 
The typical scale of the elastic cross section is
$\sigma_e^\chi\lesssim\mathcal{O}(10^{-42})~\mathrm{cm}^2$ while the
current strongest upper bound given by XENON10 and XENON100 is
$\sigma_e^\mathrm{exp}=\mathcal{O}(10^{-38})~\mathrm{cm}^2$ at
$\mathcal{O}(0.1)~\mathrm{GeV}$ of dark matter mass~\cite{Essig:2017kqs}. 
Thus no substantial constraint is imposed from the scattering with electron. 

For the lighter dark matter $a$, the elastic scattering cross section
with a proton is very small since it is suppressed by the small reduced mass
$\left(m_e^{-1}+m_a^{-1}\right)^{-1}$. 
Note that $Z^\prime$ does not mediate for the lighter dark matter $a$
unlike the $\chi$ dark matter. 
For the scattering with electron mediated by the Higgs bosons, the elastic
scattering cross section is computed as 
\begin{equation}
\sigma_e^a=
\frac{m_e^4}{4\pi(m_e+m_a)^2}\left(
\frac{\mu_{Haa}\sin\alpha}{m_H^2\langle\Phi\rangle}
-\frac{\mu_{haa}\cos\alpha}{m_h^2\langle\Phi\rangle}
\right)^2,
\end{equation}
where $\mu_{Haa}$ and $\mu_{haa}$ are given in Eq.~(\ref{eq:Haa}) and
\begin{equation}
\mu_{haa}=\sqrt{2}\lambda_{\Phi
 S}\langle\Phi\rangle\cos\alpha+\sqrt{2}\lambda_{\Sigma
 S}\langle\Sigma\rangle\sin\alpha
 -\frac{\kappa}{\sqrt{2}}\sin\alpha.
\end{equation}
The magnitude of the cross section in our setup is roughly given by
$\sigma_e^a\lesssim\mathcal{O}(10^{-43})~\mathrm{cm}^2$ which 
is small enough compared to the current experimental bound~\cite{Essig:2017kqs}.

\subsection{Cosmological Observations}
\subsubsection{Cosmic-ray, CMB and BBN Observations}
In this model with our setup, the relevant annihilation channels to
the cosmological observations are
$\chi^\dag\chi\to HZ^\prime$ and $Z^\prime Z^\prime$. 
The channel $\chi^\dag\chi\to HH$ is sufficiently suppressed by the
assumption ($\lambda_{\Phi\chi},\lambda_{\Sigma\chi}\ll1$), and 
the channel $aa\to HH$ is kinematically forbidden in non-relativistic
case since the momentum of dark matter is too small.

If $H$ or $Z^\prime$ is produced by dark matter annihilations in the
current times, 
these processes are constrained by gamma-ray and cosmic-ray observations
since the produced particles $H$ and $Z^\prime$ decay into charged particles. 
The dominant $H$ decay channel is $H\to f\overline{f}$ whose decay width is
given by the first term of Eq.~(\ref{eq:H_decay}).
For $Z^\prime$ decay, the channel $Z^\prime\to as$ is dominant, and the
decay width is given by the second term of Eq.~(\ref{eq:zp_decay}). Then $s$
subsequently decays via $s\to ae^+e^-$, and gamma-rays are produced by
Bremsstrahlung process and thus a constraint is imposed on the model~\cite{Essig:2013goa}. 

In addition, after freeze-out of dark matter particles, at so-called
dark ages, CMB is sensitively distorted by such
non-standard production of charged particles and gamma-rays.
As a result, it gives an upper bound on annihilation cross sections.
Since the upper bound of annihilation cross sections depends on the
energy spectrum of the produced $e^+e^-$, we take the strongest one as a
conservative bound, which is roughly given by
$\sigma{v}_{\chi^\dag\chi\to Z^\prime 
Z^\prime}f_\chi^2\lesssim (m_\chi/\mathrm{GeV})\times10^{-27}~\mathrm{cm^3/s}$
for $1~\mathrm{MeV}\lesssim 
m_{\chi}\lesssim 10~\mathrm{GeV}$~\cite{Slatyer:2015jla, Liu:2016cnk}. 
Thus one can see that the upper bound becomes smaller than  
the value required for the correct relic abundance ($\sim3\times10^{-26}~\mathrm{cm^3/s}$) when
$m_\chi\lesssim30~\mathrm{GeV}$ and $f_\chi=1$. 
In non-relativistic limit, the annihilation cross section for the
channel $\chi^\dag\chi\to Z^\prime Z^\prime$ is computed as
\begin{equation}
\sigma{v}_{\chi^\dag\chi\to Z^\prime Z^\prime}=
\frac{Q_\chi^4g_D^4}{16\pi m_\chi^2}\sqrt{1-\frac{m_{Z^\prime}^2}{m_\chi^2}}
\frac{16m_\chi^4-16m_\chi^2m_{Z^\prime}^2+3m_{Z^\prime}^4}{(2m_\chi^2-m_{Z^\prime}^2)^2},
\label{eq:ann_zpzp}
\end{equation}
which is velocity independent ($s$-wave). 
Therefore assuming $m_{Z^\prime}\ll m_\chi$ in
Eq.~(\ref{eq:ann_zpzp}), one can translate it into an upper bound on the gauge coupling
\begin{equation}
\left|Q_\chi g_D\right|\lesssim5.7\times10^{-3}\left(\frac{m_\chi}{\mathrm{GeV}}\right)^{3/4}f_\chi^{-1/2},
\end{equation}
for our case. This bound implies that a milli-charge $Q_\chi$ for $\chi$
is necessary when $g_D=\mathcal{O}(1)$.
This bound is shown as the green dot-dashed line in the right plot of
Fig.~\ref{fig:dd_exp}.

For the channel $\chi^\dag\chi\to HZ^\prime$, the relevant diagram is
only the left one in the third line of Fig.~\ref{fig:chi_ann}, and the
annihilation cross section in non-relativistic limit is computed as 
\begin{equation}
\sigma{v}_{\chi^\dag\chi\to HZ^\prime}=\frac{Q_\chi^2g_D^4\cos^2\alpha}{4\pi}
\frac{m_{Z^\prime}^2v^2}{(4m_\chi^2-m_{Z^\prime}^2)^2+m_{Z^\prime}^2\Gamma_{Z^\prime}^2}
\left(1+\frac{1}{3}\frac{m_\chi^2}{m_{Z^\prime}^2}\right).
\label{eq:chi_ann}
\end{equation}
As one can find from Eq.~(\ref{eq:chi_ann}), the annihilation cross section is
velocity suppressed ($p$-wave). Thus this channel is sufficiently suppressed to
evade the constraint of the cosmological observations. 

Another annihilation channel $aa\to\gamma\gamma$ induced
by the Higgs mixing also relevant to the constraints coming from the
observations of gamma-rays and CMB. The current upper bound on the
annihilation cross section is given by
$\langle\sigma{v}\rangle_{\gamma\gamma}\lesssim10^{-30}~\mathrm{cm^3/s}$
for $10~\mathrm{MeV}\lesssim m_a\lesssim100~\mathrm{MeV}$~\cite{Boddy:2015efa, Bartels:2017dpb}. 
Although the annihilation cross section for this process sufficiently
small compared to the current upper bound in most of parameter
space if $\sin\alpha\lesssim10^{-2}$ in our model, the cross
section is enhanced in particular when the second Higgs mass is close to 
the resonance ($m_H\approx 2m_a$) as we will see benchmark parameter
sets in the following section. As a result, it gives a constraint on the
Higgs mixing. In
order to evade it, the mixing has been chosen to be $\sin\alpha=10^{-3}$ in
our analysis as mentioned earlier.

The successful Big Bang Nucleosynthesis (BBN) and the effective number
of neutrinos ($N_\mathrm{eff}$) are affected if freeze-out of dark matter
occurs below $T\sim1~\mathrm{MeV}$ scale.
The constraints of the BBN and $N_\mathrm{eff}$ depend on annihilation channels.
We impose the requirement of the freeze-out temperature
$T_f\geq1~\mathrm{MeV}$ as a conservative bound~\cite{Boehm:2013jpa}.
In our case, the annihilation channels $aa\to HH\to e^+e^+e^-e^-$ and
$aa\to ss\to aae^+e^+e^-e^-$ would be relevant for the BBN and
$N_\mathrm{eff}$ constraints. 
Since the freeze-out temperature of the channels $aa\to HH$, $ss$ are
roughly given by $T_f\sim m_a/20$, this implies that
$m_a\gtrsim20~\mathrm{MeV}$ is imposed not to affect the successful BBN
and the effective number of neutrinos.

\subsubsection{Bullet Cluster}
\begin{figure}[t]
\begin{center}
\includegraphics[scale=0.75]{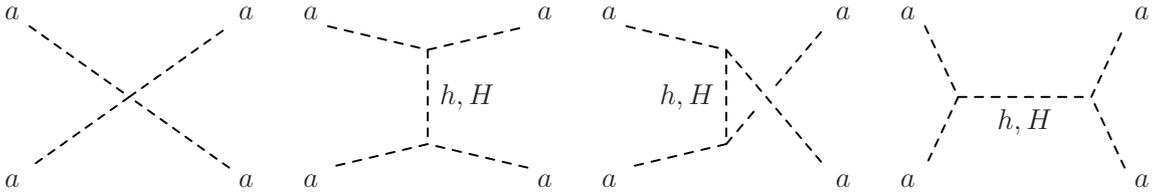}
\caption{Diagrams for self-interacting cross section.}
\label{fig:self}
\end{center}
\end{figure}

The self-interacting cross section for the dark matter $a$ is computed
from the diagrams in Fig.~\ref{fig:self} as
\begin{equation}
\sigma_\mathrm{self}=
\frac{1}{128\pi
m_a^2}\left|\frac{3}{2}\lambda_S-\frac{2\mu_{Haa}^2}{m_H^2}
+\frac{\mu_{Haa}^2}{4m_a^2-m_H^2+im_H\Gamma_H}\right|^2,
\end{equation}
where the contribution mediated by $h$ is neglected. 
The self-interacting cross section for the heavier dark matter $\chi$ is
much small and negligible even if the self-coupling $\lambda_\chi$ is $\mathcal{O}(1)$. 
The effective self-interacting cross section defined by
$\sigma_\mathrm{self}^\mathrm{eff}\equiv\sigma_\mathrm{self}f_a^2$
is bounded by the Bullet cluster observation which is given by
$\sigma_\mathrm{self}^\mathrm{eff}/m_a\lesssim1~\mathrm{cm^2/g}$~\cite{Randall:2007ph}
where $f_a\equiv \Omega_a/(\Omega_a+\Omega_\chi)$ is the fraction of
the lighter dark matter $a$. 

On the other hand, the small scale structure problems can be improved
with a large self-interacting cross section. 
The required magnitude of the
effective self-interacting cross section is roughly
$0.1\lesssim\sigma_\mathrm{self}^\mathrm{eff}/m_a\lesssim1~\mathrm{cm^2/g}$~\cite{Elbert:2014bma,
Tulin:2017ara}.

\section{Numerical Analysis}
\label{sec:5}
\subsection{Parameter scan}
The following parameter range is considered for numerical analysis:
\begin{eqnarray*}
10~\mathrm{MeV}\leq m_{a} \leq 1~\mathrm{GeV},\quad
m_{a}\leq m_{s} \leq 10m_{a},\quad
m_a\leq m_H\leq 2m_a,\hspace{0.6cm}\\
m_a+m_s \leq m_{Z^\prime} \leq 3m_s,\quad
m_{Z^\prime} \leq m_\chi \leq 10~\mathrm{GeV},\quad
10^{-3} \leq \lambda_{S\chi},\lambda_{\Sigma S} \leq 1.
\end{eqnarray*}
The other relevant parameters are fixed to be $g_D=1$,
$\epsilon_\gamma=5\times10^{-4}$, $Q_\chi=10^{-5}$ and $\sin\alpha=10^{-3}$. 
The mass of the lighter component of dark matter $a$ should be in the above
range in order to induce a large self-interacting cross section for the
small scale structure problems. 
The second extra Higgs boson mass should be in the above range for
reproducing the relic abundance of dark matter $a$.
The hidden gauge boson $Z^\prime$ should not be much heavier than $s$ so
that the decay length for the process $s\to ae^+e^-$ mediated by $Z^\prime$ becomes shorter
than a detector size in order to see the multi-Cherenkov ring events. 
For the mass of dark matter $\chi$, the mass range
$m_\chi> 10~\mathrm{GeV}$ is not considered here since the constraint of
dark matter direct detection becomes much stronger. 
The quartic coupling $\lambda_{S\chi}$ is a relevant parameter to the annihilation
process $\chi^\dag\chi\to aa$, and $\lambda_{\Sigma S}$ is relevant to
the self-interacting cross section. 

The decay length of $s$ given by $L_D=\beta_s\gamma_s\tau_s$ should be
shorter than the detector length so that the multi-Cherenkov ring events are observed
as a characteristic signature of the non-minimal dark sector.
We take a benchmark of the detector length as the diameter of Hyper-Kamiokande which is
$74~\mathrm{m}$. 
Although the momentum of the excited state $s$ produced by the inelastic
scattering $ae^-\to se^-$ has a distribution, we assume the excited state $s$ is
produced with averaged momentum for each parameter set to make our discussion simple.

\begin{figure}[t]
\begin{center}
\includegraphics[scale=0.65]{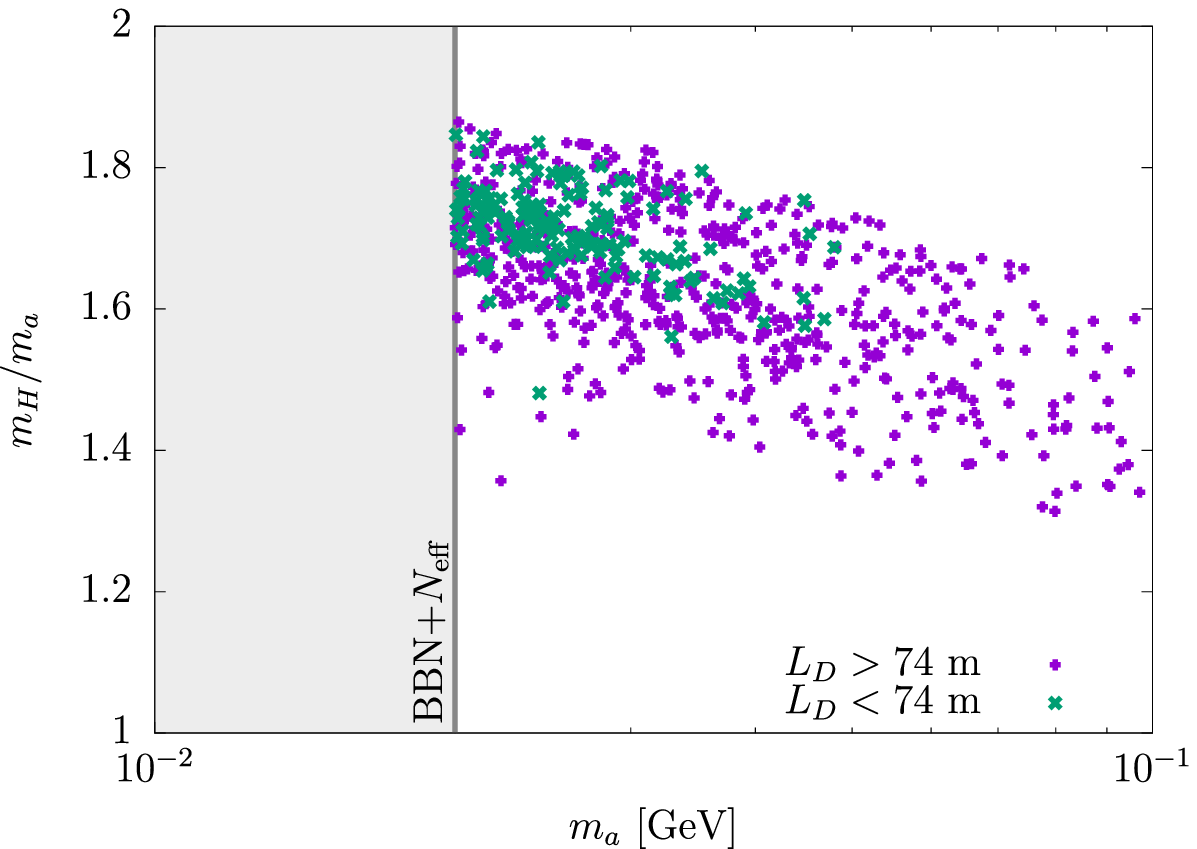}
\includegraphics[scale=0.65]{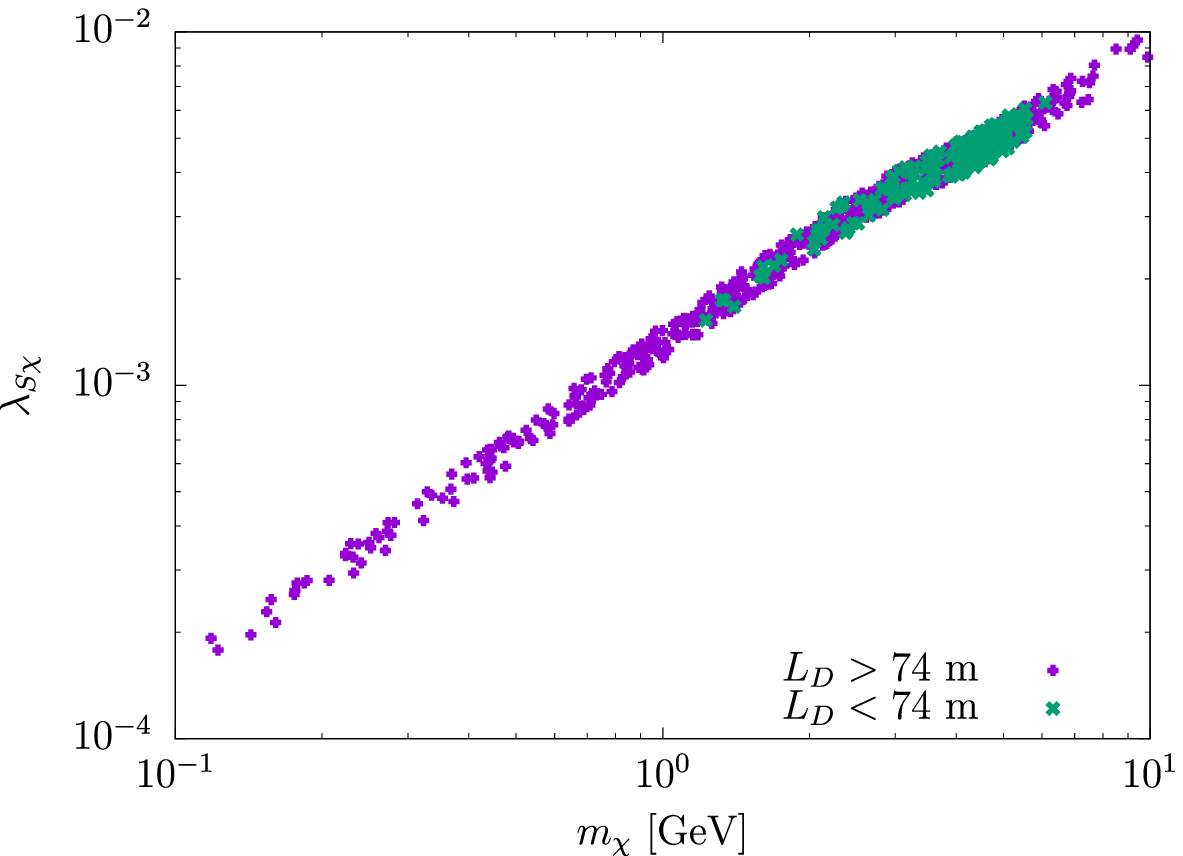}\\
\includegraphics[scale=0.65]{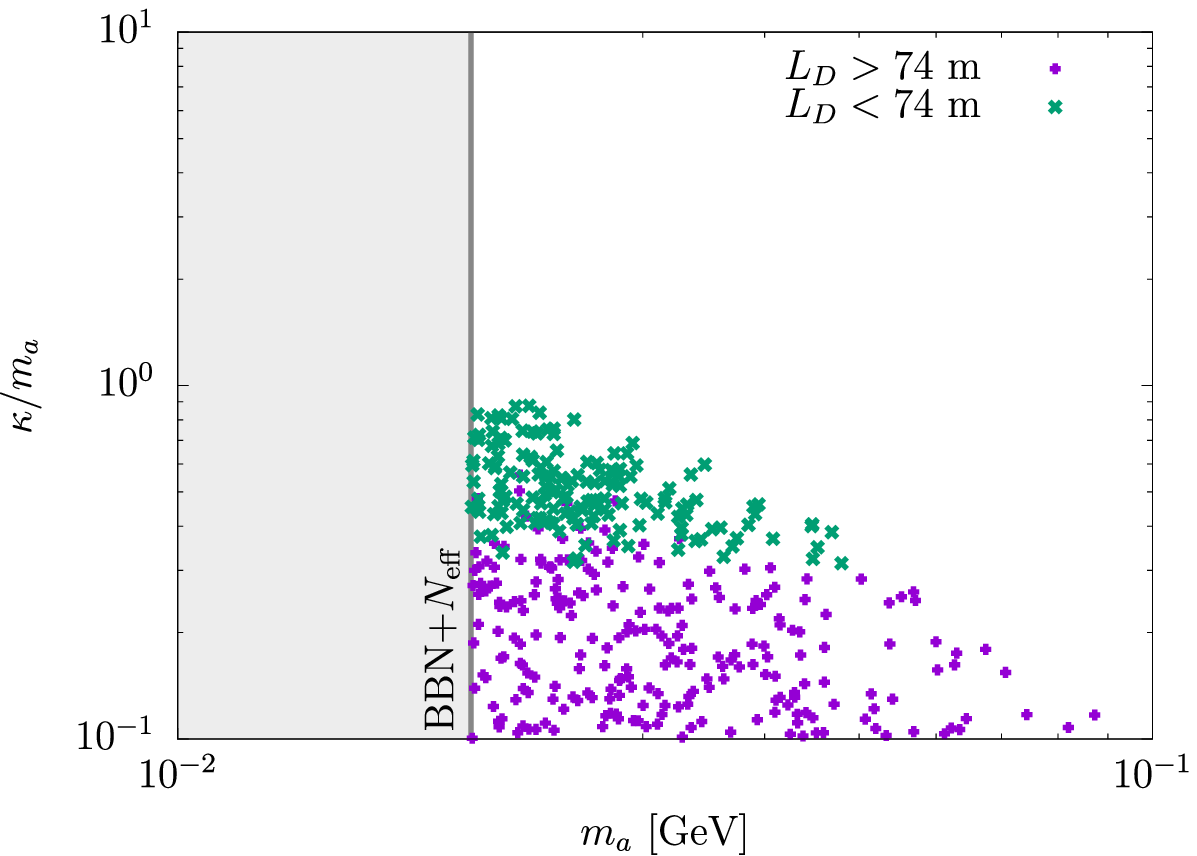}
\includegraphics[scale=0.65]{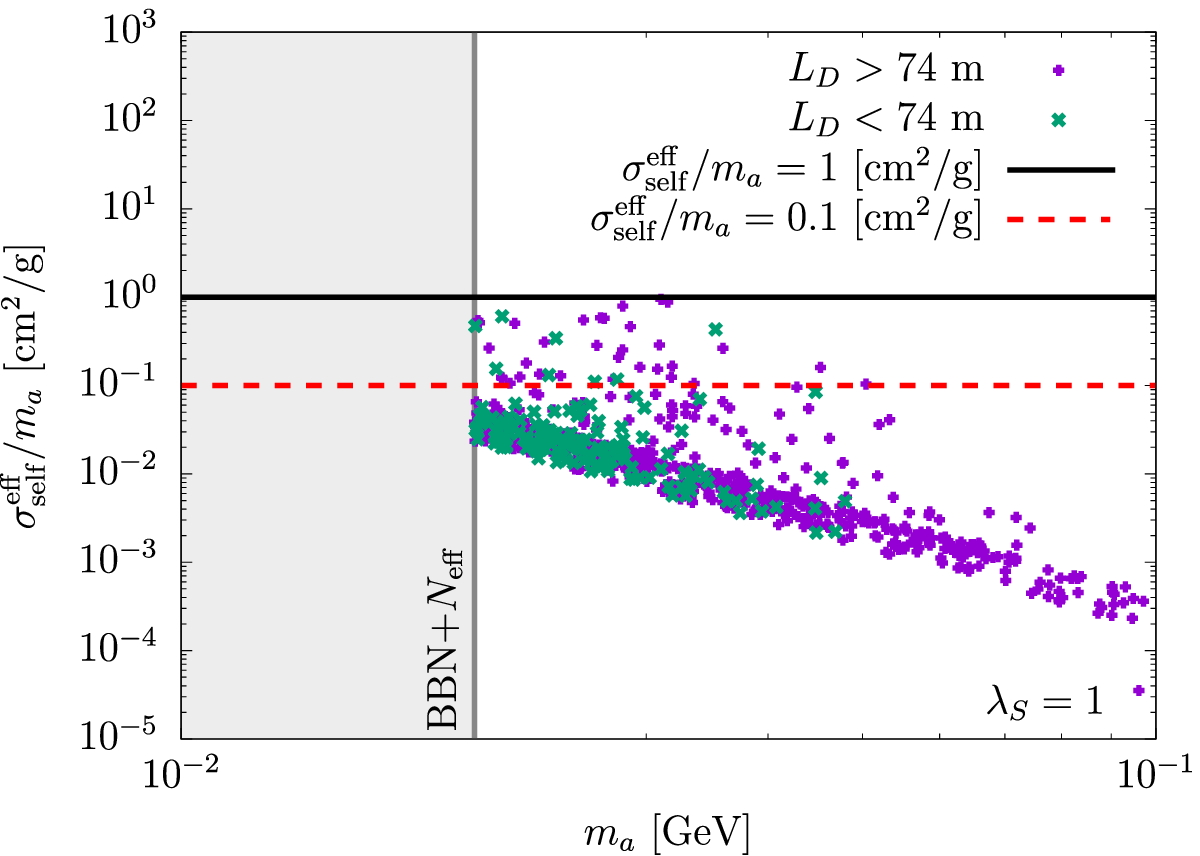}
\caption{Parameter space allowed by all the constraints where
 $g_D=1$, $\epsilon_\gamma=5\times10^{-4}$, $Q_\chi=10^{-5}$ and $\sin\alpha=10^{-3}$} 
\label{fig:num1}
\end{center} 
\end{figure}

\begin{figure}[t]
\begin{center}
\includegraphics[scale=0.65]{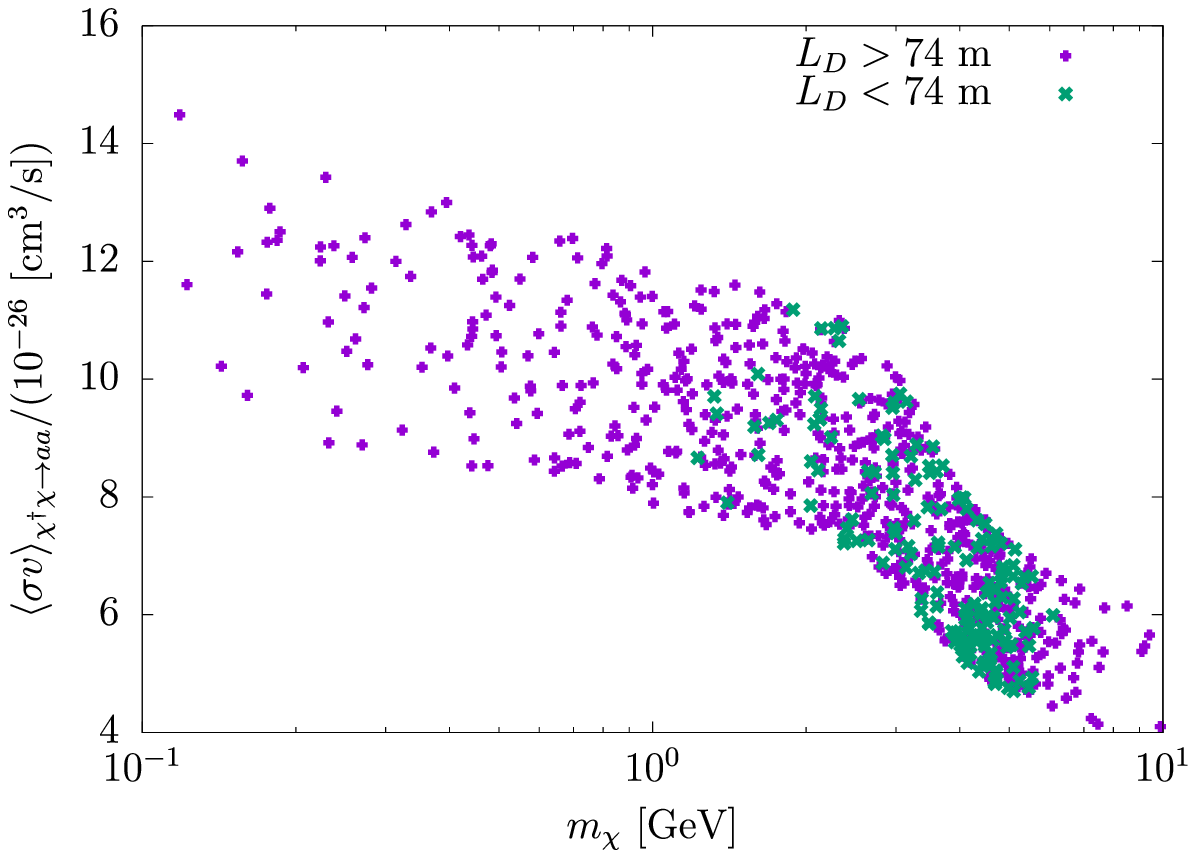}
\includegraphics[scale=0.65]{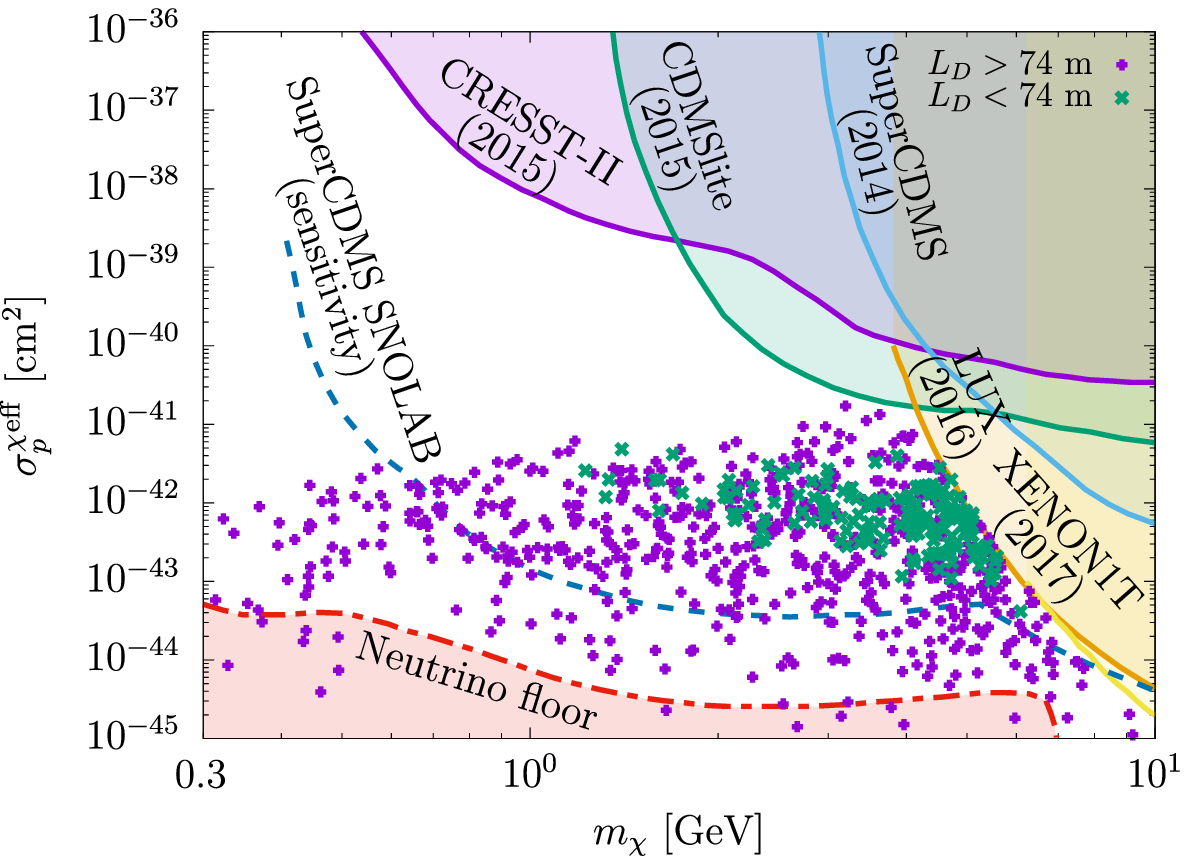}
\caption{(Left): annihilation cross section boosting the dark matter
 $a$. (Right): effective elastic scattering cross section for direct detection
 of the dark 
 matter $\chi$ where $g_D=1$, $\epsilon_\gamma=5\times10^{-4}$, $Q_\chi=10^{-5}$
 and $\sin\alpha=10^{-3}$. The same parameter sets are used with Fig.~\ref{fig:num1}.}
\label{fig:num3}
\end{center} 
\end{figure}

The parameter space consistent with all the constraints considered in
the previous section (dark matter relic
abundance, direct detection, cosmic-ray and cosmological observations)
is shown in Fig.~\ref{fig:num1}.
For the dark matter relic abundance, we take $3\sigma$ range of the
value observed by the PLANCK Collaboration
$\Omega_{\chi}h^2+\Omega_ah^2=0.1197\pm0.0022$~\cite{Ade:2015xua}.
The left upper plot in Fig.~\ref{fig:num1} shows the parameter space in
the plane ($m_a$, $m_H/m_a$) where the mass range $m_a\leq
20~\mathrm{MeV}$ is excluded by the constraints of the BBN and $N_\mathrm{eff}$.
The purple (green) points represent the allowed parameter points with
the decay length of $s$ longer (shorter) than the detector length of
Hyper-Kamiokande ($\sim74~\mathrm{m}$). 
One can see that $m_H$ is in the range
$1.3\lesssim m_H/m_a\lesssim1.9$ to reproduce the correct range of the
relic abundance $\Omega_a$ ($40\%\sim60\%$ of the total abundance) via
the forbidden channel $aa\to HH$. 
The right upper plot shows the parameter space in the plane ($m_\chi$,
$\lambda_{S\chi}$).
The relic abundance of $\chi$ is almost determined by the quartic
coupling $\lambda_{S\chi}$, and the coupling should be in the range of
$10^{-4}\lesssim\lambda_{S\chi}\lesssim10^{-2}$ to be consistent with
the $\chi$ relic abundance. Furthermore, if the requirement of the decay
length $L_D<74~\mathrm{m}$ is imposed, the parameters should be in the
range of $1~\mathrm{GeV}\lesssim m_\chi\lesssim 7~\mathrm{GeV}$ and
$10^{-3}\lesssim \lambda_{S\chi}\lesssim10^{-2}$.
The left lower plot shows the parameter space in the plane of ($m_a$,
$\kappa/m_a$). One can see that a larger $\kappa$ is required 
so that the decay length $L_D$ can be shorter where the parameter
$\kappa$ is relevant to the mass splitting between $s$ and $a$ (see
Eqs.~(\ref{eq:smass}) and (\ref{eq:amass})). 
The right lower plot shows the effective self-interacting cross section
as a function of $m_a$ where $\lambda_S=1$.
The large self-interacting cross section improving the small scale structure
problems
($0.1~\mathrm{cm^2/g}\leq\sigma_\mathrm{self}^\mathrm{eff}/m_a\leq1~\mathrm{cm^2/g}$)
can be obtained when the dark
matter mass is $m_a\lesssim50~\mathrm{MeV}$. 

With these allowed parameter sets, the annihilation cross section
boosting the lighter dark matter $a$ is shown as a function of $m_\chi$
in the left panel of Fig.~\ref{fig:num3}.
The order of the magnitude spreads in the range 
$4\times10^{-26}~\mathrm{cm^3/s}\lesssim\langle\sigma{v}\rangle_{\chi^\dag\chi\to
aa}\lesssim1.5\times10^{-25}~\mathrm{cm^3/s}$, 
and one can see that this process is dominant to determine the relic
abundance of the heavier dark matter $\chi$. 
In the right panel of Fig.~\ref{fig:num3}, the effective elastic cross section
${\sigma_p^\chi}^\mathrm{eff}$ for direct detection of dark matter is
shown as a function of $m_\chi$.
All the parameter sets with $L_D<74~\mathrm{m}$ can be explored by the
future direct detection experiment SuperCDMS
SNOLAB~\cite{Agnese:2016cpb} even though the hidden $U(1)_D$ charge is
small as $Q_\chi=10^{-5}$. 

\begin{figure}[t]
\begin{center}
\includegraphics[scale=0.65]{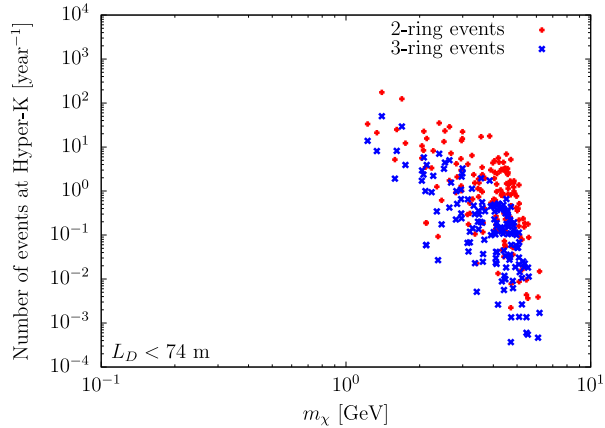}
\caption{Expected number of the multi-Cherenkov ring events at
 Hyper-Kamiokande where $g_D=1$, $\epsilon_\gamma=5\times10^{-4}$,
 $Q_\chi=10^{-5}$ and $\sin\alpha=10^{-3}$.}
\label{fig:num4}
\end{center} 
\end{figure}

With the parameter sets satisfying $L_D<74~\mathrm{m}$, the number of
multi-Cherenkov ring events at Hyper-Kamiokande is plotted as a function
of $m_\chi$ in Fig.~\ref{fig:num4} where the experimental energy
threshold $E_e,E_e^\prime,E_{\overline{e}}>0.1~\mathrm{GeV}$ is imposed.
The blue points correspond to the number of 3-Cherenkov ring events 
with the experimental angular threshold
$\theta_{e\overline{e}}\geq3^\circ$ while the red points correspond to the
number of 2-Cherenkov ring events with $\theta_{e\overline{e}}<3^\circ$. 
One can see from Fig.~\ref{fig:num4} that $\mathcal{O}(100)$
multi-Cherenkov ring events per year can be expected at most.
However if one requires a large self-interaction for the small scale structure
problems, the expected number of events would decrease.

\subsection{Benchmark Points}
\begin{figure}[t]
\begin{center}
\includegraphics[scale=0.65]{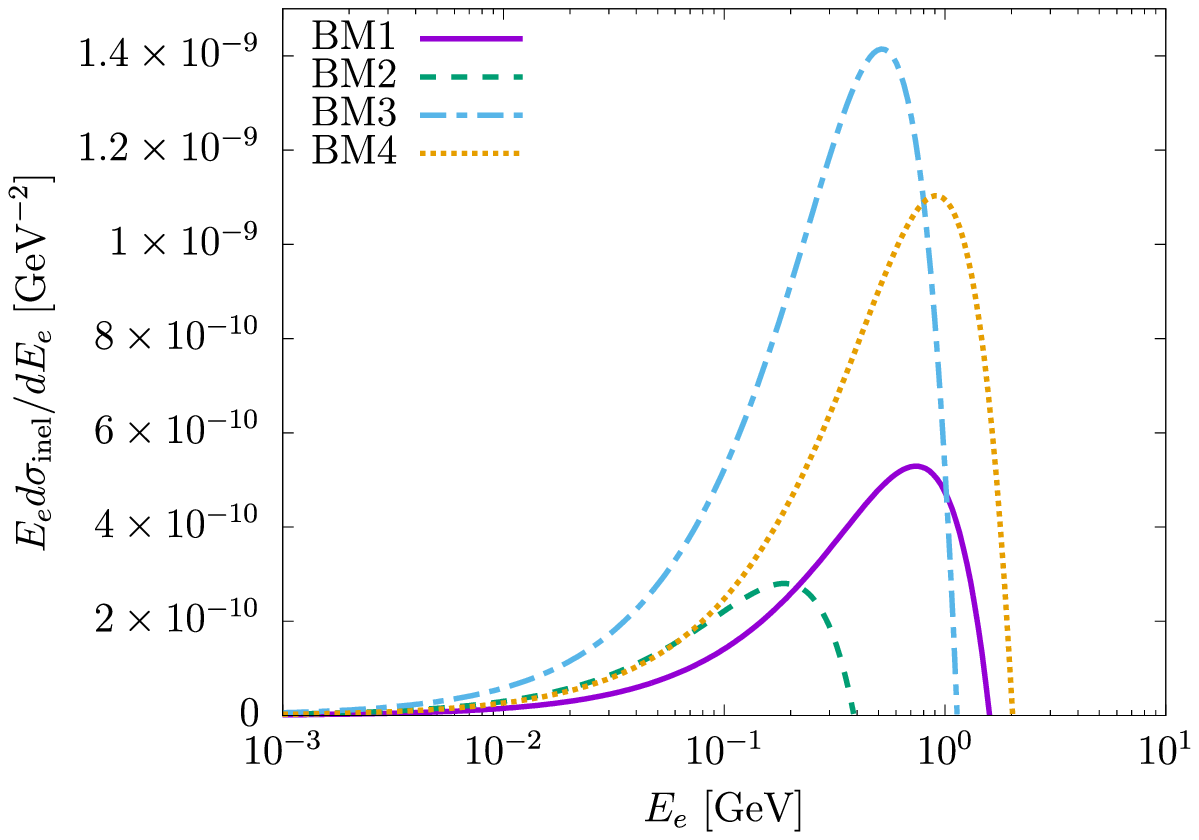}
\includegraphics[scale=0.65]{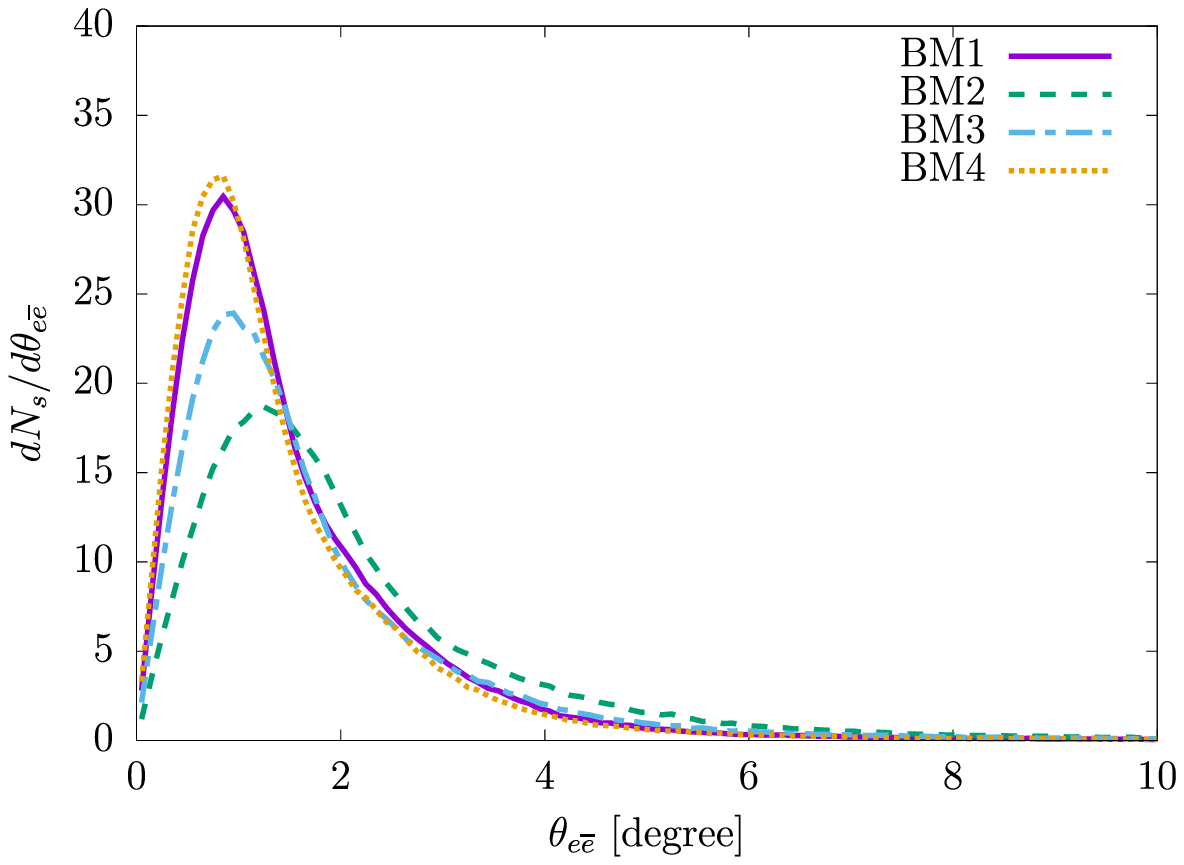}
 \caption{Energy and angular distributions for BM1-BM4.}
 \label{fig:benchmark}
\end{center}
\end{figure}

We choose some benchmark (BM) parameter sets from the above analysis as shown
in Tab.~\ref{tab:benchmark1}.
One can see that the scale of the $Z^\prime$ mass is
$m_{Z^\prime}\lesssim100~\mathrm{MeV}$, and the coupling
$\lambda_{\Sigma S}$ is $\lambda_{\Sigma S}=\mathcal{O}(0.1)$. 
With these parameter sets, the predicted physical quantities such as
relic abundance and cross sections are
summarized in Tab.~\ref{tab:benchmark2}. 
These parameter sets can improve the small scale structure problems with
$0.1~\mathrm{cm^2/g}<\sigma_\mathrm{self}^\mathrm{eff}/m_a<1~\mathrm{cm^2/g}$. 
In the left column of Tab.~\ref{tab:benchmark3}, the numbers of the total
inelastic scattering events without any cuts are listed.
For each parameter set, the energy and angular distributions
of the electrons are shown in Fig.~\ref{fig:benchmark} where the left
is $E_ed\sigma_\mathrm{inel}/dE_e$ and the right is
$dN_s/d\theta_{e\overline{e}}$. 
One can see that the inelastic cross section $\sigma_\mathrm{inel}$ is
relatively small for BM2 (left plot),
however the total expected number of events in Tab.~\ref{tab:benchmark3}
is relatively large due to the small $\chi$ mass
(Tab.~\ref{tab:benchmark1}) and the large 
annihilation cross section $\langle\sigma{v}\rangle_{\chi^{\dag}\chi\to
aa}$ (Tab.~\ref{tab:benchmark2}). 
In the centre and right columns of Tab.~\ref{tab:benchmark3},
the number of 2-Cherenkov ring events 
($E_e,E_e^\prime,E_{\overline{e}}>0.1~\mathrm{GeV}$ and
$\theta_{e\overline{e}}<3^\circ$) and 3-Cherenkov ring events
($E_e,E_e^\prime,E_{\overline{e}}>0.1~\mathrm{GeV}$ and
$\theta_{e\overline{e}}\geq3^\circ$) are summarized.
In particular, a few 3-Cherenkov ring events per year can be expected
for BM2 and BM3.\footnote{These numbers of events have been
estimated assuming the NFW dark matter profile, and would increase with
a few factor if a more cusp profile is considered such as Einasto
profile~\cite{Cirelli:2010xx}.} 

\begin{table}[t]
\begin{center}
 \caption{Benchmark parameter sets where $g_D=1$,
 $\epsilon_\gamma=5\times10^{-4}$, $\sin\alpha=10^{-3}$ and
 $Q_\chi=10^{-5}$.}
{\small
\begin{tabular}{|c||c|c|c|c|c|c|c|c|}\hline
 & $m_{a}$ [MeV] & $m_{s}$ [MeV] & $m_\chi$ [GeV] &
 $m_{Z^\prime}$ [MeV] & $m_H$ [MeV] &
 $\lambda_{S\chi}$ & $\lambda_{\Sigma S}$\\\hhline{|=#=|=|=|=|=|=|=|} 
BM1 & $21.0$ & $60.3$ & $4.86$ & $102$ & $38.4$ &
				 $5.35\times10^{-3}$ & $0.251$\\\hline
BM2 & $24.2$ & $49.1$ & $2.27$ & $75.3$ & $44.5$ &
				 $3.17\times10^{-3}$ & $0.516$\\\hline
BM3 & $28.0$ & $48.8$ & $2.99$ & $78.3$ & $50.5$ &
				 $3.34\times10^{-3}$ & $0.609$\\\hline
BM4 & $23.8$ & $54.3$ & $4.60$ & $96.4$ & $43.1$ &
				 $4.42\times10^{-3}$ & $0.386$\\\hline
\end{tabular}
}
\label{tab:benchmark1}
\end{center}
\end{table}

\begin{table}[t]
\begin{center}
\caption{Prediction with the benchmark parameter sets in Tab.~\ref{tab:benchmark1}.}
{\small
\begin{tabular}{|c||c|c|c|c|c|c|}\hline
 & $\Omega_a h^2$ : $\Omega_\chi h^2$ &
 ${\sigma_p^\chi}^\mathrm{eff}~[\mathrm{cm}^2]$ & $L_D~[\mathrm{m}]$ & 
 $\langle\sigma{v}\rangle_{\chi^\dag\chi\to aa}~[\mathrm{cm^3/s}]$ &
 $\sigma_\mathrm{self}^\mathrm{eff}/m_a~[\mathrm{cm^2/g}]$\\\hhline{|=#=|=|=|=|=|} 
BM1 & $59.0\%$ : $41.0\%$ & $6.43\times10^{-43}$ & $17.9$ &
		     $6.75\times10^{-26}$ & $0.155$\\\hline
BM2 & $58.4\%$ : $41.6\%$ & $1.62\times10^{-42}$ & $22.4$ &
		     $1.09\times10^{-25}$ & $0.344$\\\hline
BM3 & $45.4\%$ : $54.6\%$ & $2.10\times10^{-42}$ & $67.0$ &
		     $7.11\times10^{-26}$ & $0.117$\\\hline
BM4 & $44.8\%$ : $55.2\%$ & $1.10\times10^{-42}$ & $41.1$ &
		     $5.15\times10^{-26}$ & $0.132$\\\hline
\end{tabular} 
}
\label{tab:benchmark2}
\end{center}
\end{table}

\begin{table}[t]
\begin{center}
 \caption{Expected number of the multi-Cherenkov ring events at
 Hyper-Kamiokande for the benchmark parameter sets in
 Tab.~\ref{tab:benchmark1}.}
 {\small
 \begin{tabular}{|c||c||c|c|}\hline
 & Total events$~[\mathrm{year}^{-1}]$ & 2-ring events$~[\mathrm{year}^{-1}]$ & 3-ring events$~[\mathrm{year}^{-1}]$\\\hhline{|=#=#=|=|} 
BM1 & $1.61$ & $0.98$ & $0.113$\\\hline
BM2 & $21.9$ & $8.21$ & $2.21$\\\hline
BM3 & $28.6$ & $15.5$ & $2.52$\\\hline
BM4 & $4.20$ & $2.64$ & $0.266$\\\hline
 \end{tabular} 
 }
\label{tab:benchmark3}
\end{center}
\end{table}

%%%%%%%%%%%%%%%%%%%%%%%%%%%%
\section{Summary and Conclusions}
\label{sec:6}

We have constructed a model extended by three new scalar fields with a hidden
$U(1)_D$ gauge symmetry where one of the new scalar has a milli-charge of $U(1)_D$ symmetry. 
The residual symmetry $\mathbb{Z}_2\times\mathbb{Z}_2^\prime$ remains
after the spontaneous breaking of the hidden $U(1)_D$ gauge symmetry,
which guarantees two stable particles in the model. Thus the model has
two-component dark matter. 

In this model, characteristic signatures of the boosted dark
matter are induced. 
Namely, the multi-Cherenkov ring events are expected to be observed at large
neutrino detectors such as Super-Kamiokande and Hyper-Kamiokande.
Taking into account the relevant constraints such as dark matter
relic abundance, direct detection, cosmic-ray and cosmological
observations, we have explored allowed parameter space 
in the specific mass interval from $10~\mathrm{MeV}$ to
$10~\mathrm{GeV}$.
With the allowed parameter sets, we have estimated the number of
multi-Cherenkov ring events at Hyper-Kamiokande future experiment.
Our benchmark parameter sets have shown that $\mathcal{O}(100)$
multi-Cherenkov events per year are expected at most.
Moreover, if one requires a large self-interacting cross section of the
lighter component of dark matter $a$ to improve the small scale structure
problems, 3-Cherenkov ring events decrease to a few events per year. 

The mass scale of the hidden gauge boson $Z^\prime$ has been required to
be $m_{Z^\prime}\lesssim100~\mathrm{MeV}$ so that the decay length
is shorter than the detector length.
Such light $Z^\prime$ can be tested by the HPS future experiment. 
Since the $Z^\prime$ gauge boson is light enough, the elastic scattering
cross section for $\chi p^+\to\chi p^+$ has been enhanced.
As a result, the dark matter $\chi$ with
$m_\chi=\mathcal{O}(1)~\mathrm{GeV}$ can be detectable by the future dark
matter direct detection experiment SuperCDMS SNOLAB in spite of the
small charge of hidden $U(1)_D$ gauge symmetry. 

%%%%%%%%%%%%%%%%%%%%%%%%%%%%%%%%%%%

\section*{Acknowledgments}
M.~A. thank Doojin Kim, Jong-Chul Park, and Seodong Shin for early
discussions about inelastic boosted dark matter. 
M.~A. also thank Yoshinari Hayato for helpful discussions on
Super-Kamiokande experiments. 
M.~A. is supported in part by the Japan Society for the Promotion of
Sciences (JSPS) Grant-in-Aid for Scientific Research (Grant No. 25400250
and No. 16H00864). 
T.~T. acknowledges support from JSPS Fellowships for Research Abroad. 
%%%%%%%%%%%%%%%%%%%%%%%%%%%%%%%%%%%

%%%%%%%%%%%%%%%%%%%%%%%%%%%%%%%%%%%
%\newpage

\end{document}